\documentstyle[12pt]{article}
\newcommand{\beq}{\begin{equation}}
\newcommand{\eeq}{\end{equation}}
\newcommand{\bea}{\begin{eqnarray}}
\newcommand{\eea}{\end{eqnarray}}
\renewcommand{\theequation}
{\mbox{\arabic{section}.\arabic{equation}}}
\newcommand{\cL}{{\cal L}}
\newcommand{\cK}{{\cal K}}
\newcommand{\cV}{{\cal V}}

\newcommand{\oh}{\overline{h}} 
\newcommand{\T}{\textstyle}  
\newcommand{\bphi}{\mbox{\boldmath$\phi$}}
\newcommand{\bBox}{\mbox{\boldmath$\Box$}}
\input epsf
\begin{document}
\hfill \vbox{\hbox{UCLA/97/TEP/2 }
             \hbox{hep-th/9702146}}

\begin{center}{\Large\bf Superrenormalizable Gauge 
and Gravitational Theories}\footnote{Research supported in 
part by NSF Grant PHY 95-310223, and Monbusho, Japan} \\[2cm]
{\bf E. T. Tomboulis}\footnote{e-mail address: 
tombouli@physics.ucla.edu}\\
{\em Department of Physics\\
University of California, Los Angeles \\
Los Angeles, CA 90095-1547}\\[2cm]
\end{center}
\begin{center}{\Large\bf Abstract}
\end{center}
We investigate 4-dim gauge theories and gravitational 
theories with nonpolynomial actions containing an infinite 
series in covariant derivatives of the fields representing
the expansion of a transcendental entire function.
A class of entire functions is explicitly constructed such 
that: (i) the theory is perturbatively superrenormalizable; 
(ii) no (gauge-invariant) unphysical poles are introduced 
in the propagators. The nonpolynomial nature is essential; 
it is not possible to simultaneously satisfy (i) and (ii) 
with any polynomial series in derivatives. Cutting equations 
are derived verifying the absence of unphysical cuts and 
the Bogoliubov causality condition within the loop expansion. 
A generalized KL representation for the 2-point function is 
obtained exhibiting the consistency of physical positivity 
with the improved convergence of the propagators. Some physical 
effects, such as extended bound excitations in the spectrum,  
are briefly discussed.

\vfill
\pagebreak

\section{Introduction}
\setcounter{equation}{0}

In this paper we investigate 4-dimensional gauge theories 
defined by non-polynomial actions with an infinite number 
of derivatives. Specifically, we consider Langrangians of 
the form:
\[ -\frac{1}{2g^2}{\rm tr}{\bf F}_{\mu\nu}{\bf F}^{\mu\nu} 
-\frac{1}{2}{\rm tr}{\bf F}_{\mu\nu}h(-\frac{{\bf D}^2}
{\Lambda^2}){\bf F}^{\mu\nu} \]
where $h$ is a {\it transcendental entire} function having an 
infinite series expansion in the covariant D'Alembertian 
${\bf D^2}$, and $\Lambda$ some scale. Similarly, we consider 
gravitational theories with actions including terms of the 
form 
\[ R_{\mu\nu}h_2(-\frac{\nabla^2}{\Lambda^2})R^{\mu\nu} \qquad 
\mbox{and}\qquad Rh_0(-\frac{\nabla^2}{\Lambda^2})R \]
where $h_2$ and $h_0$ are entire functions.\footnote{Additional 
structures involving higher than second powers of ${\bf F},\;
R$ may be included, but will not be considered in this paper.}

When the functions $h$, or $h_2, h_0$, are taken to be  
polynomials, these are 
the Lagrangians of the familiar higher-derivative (covariant 
Pauli-Villars) regularization of gauge theory \cite{S}. 
As it is well-known, such regularization renders gauge 
theory superrenormalizable at the expense of introducing 
massive ghosts. It is easily shown that this will always be 
the case for any polynomial $h$. Here we consider the 
question whether it is possible to choose non-polynomial 
$h$ so as to obtain good UV behavior while avoiding the 
introduction of ghosts. Somewhat surprisingly, we find that 
there is a class of transcendendal entire functions, which 
can be explicitly constructed, and give a 
superrenormalizable theory, while, at least formally 
within the perturbative loop expanssion, maintaining 
unitarity and causality. Superrenormalizability and 
unitarity appear interconnected. The requirement 
that the function $h$ be entire, thus possessing no 
singularities anywhere in the finite complex plane, is 
absolutely crucial for this to be possible.  

To avoid a potential confusion at the outset, let us stress 
that what is being considered here is {\it not} the 
expansion in derivatives (powers of momenta) of the nonlocal 
effective action resulting from integration over some of the 
fields of a local field theory. Note that such an 
effective action necessarily contains 
singularities corresponding to the thresholds for production 
of the integrated out degrees of freedom. By the same token 
it cannot define a unitary S-matrix solely in terms of the 
remaining fields appearing in it, since the integrated-out 
fields still can occur in the intermediate state cuts.   

Though no nonlocal kernels are explicitly introduced in our 
actions, the dependence of the argument of the nonpolynomial 
$h$ on derivatives does introduce an effective nonlocality. 
Actions with general nonlocal kernels are, of course, 
known to lead to problems with causality. The 
nonlocality due to transcendental entire functions with 
derivative-dependent argument is, on the other hand, of a 
rather mild sort sometimes termed 'localizable' in 
distribution theory. As we will see, many things work 
for our actions pretty much as for polynomial actions 
precisely because of the similar properties of polynomial 
and nonpolynomial entire functions. 

The idea of nonpolynomial entire Langrangians as a natural 
extension of the usual polynomial ones is certainly not 
new. Efimov, in particular, pursued 
such investigations \cite{E}, mostly in the context of 
attempts to obtain finite scalar theories, some time ago. 

In the context of relativistic particle mechanics, Kato 
\cite{K} considered actions analogous to the field theory 
actions considered here. To the usual (gauge-fixed) 
particle action $(dx^\mu/d\tau)^2$, he adds a term 
$(dx^\mu/d\tau)\:f(d/d\tau)\:(dx_\mu/d\tau)$, with $f$  
some function. To this one must add Lagrange multiplier 
terms incorporating the constraints of 
reparametrization invariance. This is 
then analogous to the BRS action (\ref{act}) below, 
with $dx^\mu/d\tau$ corresponding to ${\bf F}_{\mu\nu}$. 
For appropriate choice of meromorphic function $f$, he 
finds a class of theories which includes the open 
bosonic string. It might be that the theories considered 
here have some kind of underlying extended structure 
associated with them. In this paper, however, we study 
them as a field theory problem.
 
The contents of the paper are as follows.
In section 2 the action for the gauge theory case is 
introduced. A brief review of some features of higher 
derivative regularization provides the motivation for the 
inroduction of nonpolynomial entire functions. The 
structure of the resulting interaction vertices is quite 
complicated, and is examined in section 3, with technical 
details relegated to Appendix A. Provided that the 
function $h$ satisfies appropriate asymptotic conditions, 
detailed power counting shows that only 1-loop divergences 
occur in the perturbative loop expansion. These asymptotic 
conditions are supplemented in section 4 by the requirement 
of the absence of unphysical poles at tree level. A class 
of entire functions $h$ satisfying all the requirements 
is then explicitly constructed. After discussing the relation 
between Euclidean and Minkowski Feynman rules in section 5, 
we turn to the basic issues of unitarity and causality to 
any order in the loop expansion in section 6. The special 
nature of the vertices allows one to obtain a largest time 
equation and hence generalized Cutkosky rules, which, applied 
to physical amplitudes, give the unitarity condition equations. 
No gauge-invariant unphysical poles occur in the intermediate 
states, whereas the cancellation of longitudinal and FP ghost 
gauge dependent excitations occurs as in the standard 
gauge theories and is explicitly verified. Similarly, the 
Bogoliubov causality condition equation is shown to hold. 
Details concerning the derivations are relegated to
Appendices B and C. In section 7 we consider the 2-point 
function and obtain a generalized K\"{a}llen-Lehmann 
representation for it. This makes explicit how, in this 
type of theory, the absence of unphysical excitations can 
be consistent with the improved convergence of propagators. 
With slight modifications, the entire development 
can be repeated for gravitation, which, in fact, provides 
one of the main motivations for this study. This is done in 
section 8. 

The coupling to matter is discussed, though not in 
explicit detail, in the concluding section 9.
There is a variety of potentially rather 
interesting physical effects in these gauge theories, such as 
the appearance of bound extended excitations, due to 
the modified short distance behavior. These matters are 
also briefly discussed in section 9.

\section{Action}
\setcounter{equation}{0}
 
Consider the Lagrangian\footnote{We use standard notation: 
${\bf A}_\mu=A^a_\mu t_a,\; {\bf F}_{\mu\nu}=F_{\mu\nu}^at_a, 
\; {\bf D}_\lambda\cdot{\bf F}_{\mu\nu}= \partial_\lambda 
{\bf F}_{\mu\nu} +i[{\bf A}_\lambda,
{\bf F}_{\mu\nu}]=D_{\lambda}^{ab}F_{\mu\nu}^b t^a,
\; [t_a,t_b]=if_{abc}t_c$.}  
\beq
{\cal L} = -\frac{1}{2g^2}{\rm tr}{\bf F}_{\mu\nu}{\bf
F}^{\mu\nu} -\frac{\alpha}{2}{\rm tr}{\bf F}_{\mu\nu}h(-
\frac{{\bf D}^2}{\Lambda^2}){\bf F}^{\mu\nu} 
-\frac{1}{2\xi}f_a[A]w(-\frac{\Box}{\Lambda^2})f_a[A] 
+\overline{c}_aM_{ab}c_b\;. \label{act}
\eeq
${\bf D}^2 ={\bf D_\mu D^\mu}$ and $\Box
=\partial_\mu\partial^\mu$ denote the covariant and 
ordinary D'Alembertian, respectively. $f_a[A]$ is a 
gauge-fixing function, with $w$ a gauge-fixing weighting 
function. Note that the corresponding FP ghost term $M_{ab}c_b=\delta_cf_a[A,x]$, where 
$\delta_cf_a$ is the infinitesimal transformation of 
$f_a$ with gauge transformation parameter $c_b$,
does not depend on $w$. $h$ is a given function to be 
specified, and $\Lambda$ an arbitrary mass. The coupling 
$\alpha$ can, of course, be absorbed in the definition of 
the function $h$, but is more convenient to keep it explicit.  

(\ref{act}) is invariant under the BRS transformation: 
\beq
\delta A^a_\mu = D^{ab}_\mu c_b\epsilon\;\;,\qquad
\delta c_a  =  -\frac{1}{2}f_{abc}c_bc_c\epsilon\;\;,\qquad
\delta \overline{c}_a  = -\frac{1}{\xi}w(-
\frac{\Box}{\Lambda^2})f_a[A]\epsilon\quad.\label{BRS}
\eeq

With $f^a=\partial^\mu A_\mu^a$, and the usual rescalings 
${\bf A}\to g{\bf A}, \: \xi \to 
g^2\xi$, the bare propagator is then given by
\beq
{\bf D}_{\mu\nu ab}(k) = -\frac{i}{(2\pi)^4}\frac{\delta_{ab}}
{k^2 + i\epsilon}
\left(\frac{g_{\mu\nu} - k_\mu k_\nu/k^2}{
1+g^2 \alpha h(k^2/\Lambda^2)}
+ \xi\frac{k_\mu k_\nu/k^2}{w(k^2/\Lambda^2)}\right)\: 
.\label{prop} 
\eeq
 
Further definition of (\ref{act}) hinges on the specification 
of the function $h$.

{\it Polynomial $h$ - Higher Derivative (HD) Regularization.} \ 
In the HD regularization scheme \cite{S}, the function $h(x)$ 
is chosen to be a polynomial, $h(x) = p_n(x)$, of 
degree $n$. With the weight $w(x)$ also a polynomial, and if 
$n\geq 2$, and $deg\: w \geq n$, straightforward power 
counting shows that, at finite $\Lambda$, the only divergent 
diagrams are one-loop diagrams with $0,2,3$ and $4$ external 
gauge field legs and no external ghost legs. All 
other one-loop diagrams, and all IPI multi-loop diagrams 
are superficially convergent. Superficially 
convergent multi-loop diagrams may, of course, still 
contain subdivergences due to one-loop subdiagrams. The 
theory is thus rendered superrenormalizable. 

To completely regulate the theory then, the remaining 
one-loop divergences must be regulated separately. 
Dimensional regularization is straightforward to implement 
and very convenient for this purpose. Alternatively, and 
perhaps more in the spirit of the original HD scheme, 
additional Pauli-Villars (PV) regulators may be used  
\cite{S}. 

For discussion of renormalization, it is very convenient 
to note that by taking $deg\; w$ sufficiently high all 
gauge dependent divergences disappear; renormalization 
may then be performed by the addition of only 
gauge invariant counterterms. So at finite  $\Lambda$, 
the remaining one-loop 
divergences are formally manifestly gauge-invariant for 
$deg\; w \geq n$, since FP ghost field and vertex 
renormalizations are finite. They may be removed by adding 
the one-loop $F^2_{\mu\nu}$ counterterm; it is important 
to note that the function $h(x)=p_n(x)$ does not get 
renormalized. These statements can, of course, be made 
rigorous only in the presence of appropriate one-loop 
regularization. Dimensional regularization works  
well. The introduction of additional PV regulators, 
on the other hand, requires considerable care to 
avoid conflicts with gauge or BRS invariance, and 
has been the subject of several recent investigations; 
for a review and discussion see Ref.\cite{S1}, and references 
therein.

In HD regularization, where $h(x) = p_n(x)$, the theory 
(\ref{act}) is rendered superrenormalizable at the expense 
of introducing ghosts. Indeed, as it is evident from 
(\ref{prop}), the transverse part of the propagator 
acquires $n$ additional poles from the $n$ zeroes of the 
polynomial $1+g^2\alpha p_n(k^2/\Lambda^2)$. Note that some 
of these will, in general, be complex. The residues of 
some of these poles will necessarily be negative 
(more generally,  have a negative real part). This follows 
from the improved UV behavior of (\ref{prop}). Indeed, 
by the factorization theorem for polynomials and 
partial fraction decomposition one may write: 
\beq
\frac{1}{k^2(1 + g^2\alpha p_n(k^2))} = \frac{r_o}{k^2} + 
\sum_{i=1}^n \frac{r_i}{k^2 -M^2_i}\quad. \label{fract}
\eeq 
The assertion that at least one $r_i$ must be negative 
follows immediately by multiplying 
(\ref{fract}) by $k^2$ and taking the large $k^2$ limit. 
More generally, the spectral function in the 
K\"{a}llen-Lehmann representation for the dressed 
propagator must contain negative contributions and 
satisfy a superconvergence relation.

{\it Entire transcendendental $h$.} \ 
The question we consider in this paper then is: 
is it possible to choose the function $h(x)$ in (\ref{act}) 
so that no unphysical poles are introduced while at the 
same time maintaining the (super)renormalizability of the 
theory? 

It is, of course, clear from the above argument that 
the answer is no as long as $h(x)$ is taken to be a 
polynomial of any finite order (fundamental theorem of 
algebra!). One, therefore, has to consider non-
polynomial functions.
Now a polynomial is an entire function, i.e. holomorphic 
anywhere in the finite complex plane. This property is 
necessary for the action to be well-defined everywhere 
(including the complex domain needed for analyticity 
and unitarity considerations). The natural 
generalization of a polynomial possessing this 
property is a transcendental (i.e. 
non-polynomial) entire function, which we will take 
$h(z)$ to be. This means that it can be represented by 
an everywhere convergent power series about any point, 
in particular the origin: 
\beq
h(z) =  \sum_{n=0}^\infty \;a_n z^n \label{hexa} 
\eeq
with $a_n=\frac{1}{n!}h^{(n)}(0)$. Infinite radius of 
convergence (in fact, absolute convergence) implies $\lim_{n
\to\infty} \;\sqrt[n]{|a_n|} = 0$. 
The operator function $h(-{\bf D}^2/\Lambda^2)$
is then defined through (\ref{hexa}) as a power series 
in the covariant D'Alembertian ${\bf D}^2$, and gives a 
well-defined non-polynomial action (\ref{act}). 

Recall \cite{T} that the standard growth scale for entire 
functions is based on exponentials of powers as 
comparison functions: if $f(z)$ is of order $\rho$, 
then $\exp\,(r^{(\rho - \epsilon)}) 
< \max_{|z|=r}\:|f(z)| < \exp\,(r^{(\rho + \epsilon)})$ for 
arbitrary positive $\epsilon$, and $r$ sufficiently large. 
(Polynomials are of order zero.) It may thus at first appear 
that controllable UV behavior would not be possible. The overall 
growth scale provided by $\rho$, however, ignores any dependence 
of growth on the direction in which $z$ grows large. A more 
refined growth measure is obtained by defining\footnote{Let 
\beq
M_f(r,\alpha,\beta) \equiv \max_{\alpha\leq\theta\leq\beta}\; 
|f(re^{i\theta})| \qquad ,\label{mod}
\eeq
and define the order $\rho(\alpha,\beta)$ of $f(z)$ in the 
angle $\alpha \leq \arg z \leq \beta$ by 
\beq 
\overline{\lim_{r \to \infty}}\; \frac{\ln\,\ln\:M_f(r,
\alpha,\beta)}{\ln r} \qquad .\label{rho}
\eeq 
A concise and fairly complete account of the theory of growth 
of entire functions is given in Chapter 1 of the second 
reference in \cite{T}.} the order $\rho(\alpha,\beta)$ of 
$f(z)$ in the angle $\alpha \leq \arg z \leq \beta$. 
It is a remarkable property of entire functions that, for 
appropriate $f(z)$, $\rho(\alpha,\beta)$ may range from 
zero to arbitrarily large values as $\alpha, \beta$ vary. 
This property can be exploited in order to obtain controllable 
UV behavior \cite{E}. Our basic requirement will be that $h(z)$ 
in (\ref{act}) exhibit at most polynomial behavior along 
the real axis.

\section{Perturbative expansion and renormalization}
\setcounter{equation}{0} 

With $h$ a transcendental entire function, as in eq. 
(\ref{hexa}), the action (\ref{act}) now possesses, in 
addition to the usual YM vertices, 
an infinite set of interaction vertices.  In an 
obvious notation, suppressing spacetime and group indices 
and with\\ 
$\delta^n/\delta{\bf A}^n \equiv \delta^n/
\delta{\bf A}(x_1)\ldots\delta{\bf A}(x_n)$, an $N$-point 
$h$-dependent vertex is given by:

\bea
V^{(N)}(x; x_1,\ldots,x_N) & \equiv & \cV^{(N)}(
\{\partial_{x_i}\})\prod_{i=1}^N\delta(x-x_i) \nonumber\\
  & = &{\rm tr}\left(\left.\frac{\delta^{n^{\prime\prime}}{\bf 
F}[A]}{\delta{\bf A}^{n^{\prime\prime}}}\right|_{{\bf A}=0}\cdot
v^{(n)}(x,\partial_x; x_1,\ldots,x_n)\cdot\left.\frac{\delta^
{n^{\prime}}{\bf F}[A]}{\delta{\bf A}^{n^{\prime}}}\right|_
{{\bf A}=0}\right) \nonumber\\
  & \equiv & {\rm tr}\;{\bf F}^{(n^{\prime\prime})}\cdot v^{(n)}
\cdot{\bf F}^{(n^\prime)}, \qquad N=n^{\prime}+n+n^{\prime\prime} 
\label{V}
\eea 
where 
\bea
v^{(n)} & = & \frac{\delta^n}{\delta{\bf A}^n}\: 
\sum_{r=0}^\infty \; a_r\; \left(-\frac{{\bf D}^2[{\bf A}]}{
\Lambda^2}\right)^r_{|{\bf A}=0} \nonumber\\
& = & \sum_{l= \frac{n}{2}\;(\frac{n-1}{2}+1)\atop n\:{\rm even
 (odd)}}^n \sum_\sigma\; \sum_{r=l}^{\infty}\: a_r \; 
{\cal S}_{r,l,n}^\sigma 
\left( \left(-\frac{\bBox}{\Lambda^2}\right)^{(r-l)}, 
\left(\frac{1}{\Lambda^2}
\frac{\delta^b}{\delta{\bf A}^b}[-{\bf D}^2 + \bBox]_{|{\bf 
A}=0}\right)^l\right) \nonumber\\
  & \equiv &  \sum_{l= \frac{n}{2}\;(\frac{n-1}{2}+1)\atop n\:
{\rm even (odd)}}^n \sum_\sigma \:v^\sigma_{l,n}(x,
\partial_x;\; x_1, \ldots,x_n) \quad . \label{v} 
\eea 
In (\ref{v}), ${\cal S}^\sigma_{r,l,n}$ stands for the sum of 
all possible ways of distributing $(r-l)$ powers of $(-\bBox)$ 
in the $l+1$ positions among an ordered sequence, indexed by  
$\sigma$, of $l$ factors of $\frac{\delta^b}{\delta{\bf A}^b}
[-{\bf D}^2 +\bBox]|_{{\bf A}=0}$, $\:b=1,2$ (see A.1). 
The total number of $\delta/\delta{\bf A}$'s among these $l$ 
factors is $n$, and $b=1$ or $2$ since $[-{\bf D}^2 +\bBox]$ 
is at most bilinear in ${\bf A}$. The total number $\Gamma$ 
of such ordered sequences is then  
\beq
\Gamma = (n-l)! {l \choose (n-l)} l! = \frac{l!^2}{(2l-n)!} 
\quad .\label{oseq}
\eeq

The structure of $v^\sigma_{l,n}$ is examined in Appendix 
A, where it is explicitly reexpressed in terms of 
the function $h$ and its derivatives $h^{(m)}$.   
(It is of course important that one be able to do this,  
so that the asymptotic behavior of (\ref{V}) can be 
related to that of $h(z)$.) For arbitrary configuration of 
momenta $q_1,\ldots, q_N$ carried by the legs of the vertex 
(\ref{V}), $v^{(n)}$ is given through (\ref{Ssum}), 
(\ref{SJform})-(\ref{Cform}) as a sum of 
products of rational functions of momenta and $h$ 
or its derivatives. Our fundamental requirement is that 
$h$ behaves asymptotically for real values of its 
argument as a polynomial. The assumptions of the power 
counting theorem \cite{W} are then satisfied.\footnote{The 
vertices (\ref{V}), and hence integrands of graphs, are, 
in the terminology of reference \cite{W}, functions in the 
class $A_n$.} Let $q_i= c_i k + p_i, \: i=1,\ldots, 
M\leq N$ for some set of constants $c_i$ and fixed finite 
momenta $p_i$, and with k growing arbitrarily large.
By choosing the $c_i$'s and $M$ the growth of the vertex 
(\ref{V}) along every hyperplane in the space of the 
vertex momenta can then be examined. As shown in Appendix A, 
in all cases the leading asymptotic behavior of 
$v^\sigma_{l,n}$ is given by a sum of terms that 
grow at most either as\footnote{In the following, to 
avoid cluttering the notation 
we often write $h(k^2/\Lambda^2)=h(k^2)$} 
\beq
 h^{(s)}(k^2)\;k^{2s-n} \: \:\:,\qquad 
0\leq s \leq l \quad, \label{as1}\\
\eeq
\beq
\mbox{or}\qquad \qquad\: \frac{1}{k^2}k^{n^{{\rm int}}} \qquad
 \qquad\qquad \qquad\qquad\qquad, \label{as2}   
\eeq
where $n^{{\rm int}}$ is the number among the legs of 
$v^\sigma_{l,n}$ carrying momentum of order $k$. Let 
\beq
\gamma \equiv \lim_{|z|\to\infty \atop Im\: z \to 0} \left( 
\frac{\ln |h(z)|}{\ln|z|}\right) \quad. \label{gamma}
\eeq 
Note that finiteness of the limit $\gamma$ defined 
in (\ref{gamma}) implies the requirement that $h(z)$ 
exhibit at most polynomial behavior at infinity on the 
real axis. Detailed power counting (Appendix A) shows that 
UV divergences arise solely from terms with growth of type
(\ref{as1}) provided $\gamma \geq 2$.
Then the superficial degree of divergence of a 1PI graph 
$G$ with $L$ loops, $E$ external gauge boson lines and 
no external ghost lines is:
\beq 
\delta_G = 4 - 2\gamma (L-1) - E\quad,\qquad \gamma \geq 2 
\:\;.\label{divdeg}
\eeq
Thus only 1-loop diagrams with $E = 2, 3,$ or $4$ are 
superficially divergent. All other diagrams, i.e. 
1-loop graphs with $E > 4$, and all $L>1$ graphs with 
any number of external legs are superficially finite. 
Also, graphs with any number of external ghost 
legs are convergent for all $L$. The theory is then 
superrenormalizable by power counting, and the 1-loop 
divergences present are gauge-invariant. 
Renormalization of (\ref{act}) 
is thus performed very simply  by the addition of only 
the gauge-invariant 1-loop ${\bf F}^2$ counterterm: 
\beq
{\cal L}_R = -\frac{1}{2g^2}{\rm tr}{\bf F}_{\mu\nu}
{\bf F}^{\mu\nu} 
-\frac{\alpha}{2}{\rm tr}{\bf F}_{\mu\nu}h(-\frac{{\bf D}^2}
{\Lambda^2}){\bf F}^{\mu\nu}  -\frac{1}{2g^2}\;(Z_3 - 1)
{\rm tr}{\bf F}_{\mu\nu}{\bf F}^{\mu\nu}\quad, \label{actr}
\eeq 
where $g=g(\mu)$ is now the renormalized coupling 
fixed at some renormalization scale $\mu$. With the 
customary rescaling 
${\bf A} \to g{\bf A}\;\;,\;\; \xi \to g^2 \xi\;$, 
equations (\ref{prop}) and (\ref{V}) then again give the 
propagator and vertices, now in terms of the renormalized 
coupling. 

It is crucial for what follows that the function $h$ 
does not get renormalized, or, more precisely, {\it the 
functional dependence on its argument is not altered under
renormalization.}

We still have to show that functions $h$ with the required 
properties can be found. 

\section{Construction of the entire function $h$}
\setcounter{equation}{0}
In view of the form of the denominator in (\ref{prop}), it 
is convenient to define 
\beq
\oh(z) \equiv 1 + g^2 \alpha h(z) \quad.\label{barh}
\eeq

We require that the function $h(z)$ be an entire 
transcendental function with the following properties: 
\begin{enumerate}
\item[(i)] $\oh(z)$ is real and positive on the  
real axis, and has no zeroes anywhere in the complex 
plane, $|z| < \infty$. 
\item[(ii)] $|h(z)|$ has the same asymptotic behavior along 
the real axis at $\pm\infty$. 
\item[(iii)] There exists $\Theta > 0$ such that
\[  |h(z)| \stackrel{|z|\to\infty}{\longrightarrow} 
|z|^\gamma \:\:, \quad \gamma \geq 2 \]
for arguments in the cones: 
\beq 
{\cal C} = \left\{ z \:\:| \:\:-\Theta < \arg z < \Theta\:
 ,\:\:\pi - \Theta < \arg z < \pi +\Theta\right\}\quad,\
 \quad 0<\Theta<\pi/2  \:.\label{cones}
\eeq
\end{enumerate}

Condition (i) is the requirement that no poles appear in 
the transverse bare propagator (\ref{prop}) other than 
the physical (positive residue) massless gauge boson pole. 
Reality of the action (\ref{act}) is ensured. 
Condition (iii) ensures that
the power counting requirements for superrenormalizability 
of the previous section are satisfied. The appropriate 
asymptotic behavior is imposed in compliance with (ii), and 
not only on the real axis but in conelike regions 
surrounding it. This is in fact necessary since amplitudes 
are defined as boundary values of complex functions on 
the real axis. Condition (ii) is not strictly necessary as 
far as power counting requirements go. Rigorous 
power counting is considered to be performed in 
Euclidean space.\footnote{But it can actually also be done 
directly in Minkowski space using Zimmermann's trick \cite{Z}.}
It is, however, necessary if we are to obtain the usual 
formal identity in (asymptotic behavior of) Feynman rules for 
Euclidean and Minkowski $k^2$. This is important in 
the derivation of unitarity cutting rules. (The relation 
between Euclidean and Minkowski amplitudes is discussed in 
the following section.)    

The conditions (i) - (iii) lead directly to a general 
form of $h$. It is a basic result that an entire 
function with no zeroes anywhere in the complex plane 
can only be the exponential of an 
entire function.Thus, if (i) were to be satisfied, 
we must have: \\ 
$\oh(z) = \exp H(z)\;,\quad$ where $H$ is entire, 
and, from (iii), should  exhibit logarithmic asymptotic 
behavior in the region ${\cal C}$. Thus we arrive at 
the form:
\beq 
H(z) = \; \int_0^{p_\gamma(z)} \: 
\frac{1 - \zeta(w)}{w}\;dw 
\quad ,\label{Hform}
\eeq 
with: \\
(a) $p_\gamma(z)$ a real polynomial of degree $\gamma$, and 
    $p_\gamma(0) = 0$,\\
(b) $\zeta(z)$ entire and real on the real axis, and
    $\zeta(0) = 1$,\\
(c)
\beq
|\zeta(z)| \to 0 \quad \mbox{for} \quad 
|z| \to \infty \:\:,\quad z \in {\cal C} \quad. \label{asw}
\eeq

We take for $h$ in (\ref{act}):  
\beq
h(z) = \left( e^{H(z)} - 1\right) \quad , \label{hform}
\eeq
where $H(z)$ is given by (\ref{Hform}). Since $\zeta(z)$ 
is bounded in domains extending to infinity, it must be 
of order $\rho > 1/2$ (Wiman's theorem), and $h(z)$ 
is of infinite order.

The absence of zeros requirement in (i) is now satisfied 
if we set $g(\mu)^2 = \alpha^{-1}$. Since $\alpha$ is an 
arbitrary parameter, this is always possible. 
We should, however, consider its meaning under changes of 
renormalization scale. 

Assume that, at a given scale $\mu$, we have $g
(\mu)^2 \neq \alpha^{-1}$. Now evolve $g(\mu)$ to the value 
of $\mu=\mu_0$ where $g(\mu_0)^2 = \alpha^{-1}$. 
(We make the 'naturalness' 
assumption that, with (\ref{hform}) in (\ref{act}), one of 
the couplings $g^{-2}$ and $\alpha$ is not unnaturally small 
or large relative to the other.) By RG invariance, physical 
quantities are unchanged under this change of the 
parametrization of the theory in terms of $g(\mu),\;\mu$ 
to one in terms of $g(\mu_0),\;\mu_0$.

Another way of looking at this is to note that a 
specification of $g(\mu)=g_1$ at one value of $\mu$ is 
equivalent to a specification  
$\bar{\Lambda}=\bar{\Lambda}_1$ of a RG-invariant scale 
$\bar{\Lambda}$ corresponding to the renormalization of 
$g$. Suppose a different specification $g(\mu)=g_2$ 
is made, and  evolving from $g_2$ one obtains 
$g(\mu_0) = g_1$. Then,
by a familiar argument,  
$\bar{\Lambda}_2=\left(\frac{\mu_0}{\mu}\right)\bar{
\Lambda}_1\;$; and the scale $\Lambda$ must be rescaled by 
the same amount to keep the same numerical value. So two 
versions of the theory specified by $g(\mu)^2 = \alpha^{-1}$ 
and $g(\mu_0)^2 = \alpha^{-1}$, respectively, for given 
$\alpha$, differ only by a change in mass scale.

We may indeed always assume, with no loss of generality, 
that we work with a renormalization prescription such that 
$g(\mu_0)^2 = \alpha^{-1}$ at some convenient renormalization 
scale $\mu_0$. In fact, note that any split between a 
renormalized $1/g^2$ 
and a constant part (coefficient $a_0$ in (\ref{hexa})) in 
$h$ is a renormalization prescription. Since the $h$ 
modification to the action is relevant 
only in an UV regime set by the scale $\Lambda$, it is 
natural to fix $\mu_0$ of order $\Lambda$.

Choosing to renormalize at $\mu_0$ then, (i) is 
satisfied, and the bare transverse propagator has no 
additional poles. This will be extremely convenient in 
the following, in particular in deriving cutting rules and 
equations for unitarity and causality. It is clearly 
not essential, however, and we may choose a different 
renormalization point. The technical nuisance then 
would be that, if we want to avoid dealing 
with fictitious poles, we must work 
with cutting equations in terms of {\it dressed} 
propagators.\footnote{In general, cutting rules in terms of 
dressed propagators often become necessary in order to  
accomodate complications due to mass renormalization pole shifts, 
unstable particles etc., see \cite{V}.} 
Indeed, consider any other $\mu$ where 
(\ref{prop}) {\it will} have additional poles at $k^2$ 
such that: 
\beq
    \exp H(k^2/\Lambda^2) = (1 - \frac{1}{g(\mu)^2\alpha})
\quad. \label{zeros}
\eeq
In fact, it will have an {\it infinite} number of (complex) 
poles since (\ref{zeros}) must have an infinite number of 
roots (Picard's little theorem). They are all, however, 
clearly unphysical since their position moves with $\mu$ 
and is actually driven off to infinity as $\mu \to \mu_0$. 
We know, of course, that they must cancel, at any fixed $\mu$, 
by RG invariance. Thus, if in analogy to $\oh$, occuring in 
the transverse bare 2-point function, we consider the dressed 
counterpart 
\beq
\tilde{\oh}(z) \equiv 1 + g^2(\mu)\,\alpha h(z) + g^2(\mu)\, 
\pi (z, \Lambda/\mu, g(\mu),\mu) \quad ,\label{dbarh}
\eeq
occuring in the RG invariant transverse inverse full 2-point 
function, cp. eq. (\ref{fullp}) below\footnote{An explicit 
factor of $(2\pi)^4g^2$ has been factored out in the definition 
of the self-energy $\pi$ in (\ref{dbarh}) compared to that 
in (\ref{self}), (\ref{fullp}).}, we have: 
\bea
\frac{1}{g(\mu_0)^2} + \alpha h(k^2) + 
\pi(k^2,g(\mu_0),\mu_0) & = & 
\frac{1}{g(\mu_0)^2} + \alpha h(k^2) + 
\pi(k^2,g(\mu),\mu) + \Delta(g,\mu,\mu_0) \nonumber\\
        & = & \frac{1}{g(\mu)^2} + \alpha h(k^2) + 
                          \pi(k^2,g(\mu),\mu)\quad.
\label{RGinv}
\eea
Here $\Delta(g,\mu,\mu_0)$ is the finite difference between 
the counterterms renormalizing the self-energy at scales 
$\mu$ and $\mu_0$, respectively - it provides the finite 
renormalization between $g(\mu)$ and $g(\mu_0)$, and the 
shift responsible for the zeroes in the bare part, (eq. 
(\ref{zeros})), at scale $\mu$. The equality (\ref{RGinv}) 
shows how these zeros in the bare part at scale $\mu$ cancel 
against the self energy contribution to reproduce 
their absence at $\mu_0$. In short, the requirement (i) 
above may be replaced by the RG invariant statement that no 
zeros (\ref{zeros}) survive in $\tilde{\oh}$. But this 
is automatically satisfied once (i) is fulfilled at some 
renormalization scale $\mu=\mu_0$. From now on we will 
assume that the coupling has been renormalized at scale 
$\mu_0$. 

Clearly, many examples of functions that satisfy the stated 
conditions on $\zeta(z)$ can be given. An obvious choice is to 
assume exponential fall-off and take:
\beq
\zeta(z) = \exp - \left( C_n z^{2n} + \cdots + C_1 z^2 
\right)\qquad (C_n > 0\;, \:\: n\geq 1)\:. \label{wform}
\eeq 
Define $4n$ equal angular sectors $\omega_j^\epsilon$ with 
common vertex at the origin: 
\beq
\omega_j^\epsilon:\quad (2j - 3)\frac{\pi}{4n} + 
\frac{\epsilon}{2n} < \theta < 
(2j - 1)\frac{\pi}{4n} - \frac{\epsilon}{2n} \label{sects}
\eeq
with $0<\epsilon\ll 1$. The sectors $\omega_j \equiv 
\omega_j^{\epsilon=0}$ then divide the plane in $4n$ sectors. 
Now elementary estimates show that, for any (arbitrarily) small  
$\epsilon>0$, and  $|z|$ larger than 
some number $Z(\epsilon)$,  one has: 
\bea
|\zeta(z)| >  \exp\left[C_n|z|^{2n}(1 - 
\epsilon)\epsilon)\right] 
&\stackrel{|z|\to\infty}{\longrightarrow}& \infty \quad,
\qquad z\in 
\omega_j^\epsilon\: ,\quad j\quad\mbox{ even} \nonumber\\
|\zeta(z)| <  \exp\left[-C_n|z|^{2n}(1 - \epsilon)\epsilon)
\right] &\stackrel{|z|\to\infty}{\longrightarrow}& 0 
\quad,\qquad  z\in \omega_j^\epsilon\: 
,\quad j\quad\mbox{odd}\;. \label{assects}
\eea
The cones ${\cal C}$ in (\ref{cones}) then are the sectors 
$\omega_1$ and $\omega_{1+2n}$, with $\Theta = \pi/4n$. 
Note that as $n$ increases $\Theta$ decreases, but the total 
area of the angular sectors in which $\lim_{|z|
\to\infty}\zeta(z)\to 0$ always occupies half of the 
total plane area. 

The simplest choice $n=1$. Then $h$ is given by (\ref{hform}) 
with 
\beq
H(z) = \left[ \;\sum_{m=1}^\infty\: (-1)^{(m-1)} 
\frac{C^m}{m!}\frac{p_\gamma(z)^{2m}}{2m}\right] 
\qquad,\label{Hform1}
\eeq
and $\Theta = \pi/4$.

\section{Relation between Minkowski and Euclidean formulation}
\setcounter{equation}{0}

By construction, the function $h(z)$ exhibits polynomial 
asymptotic behavior in certain directions in the complex 
plane, in particular the cones ${\cal C}$, eq. (\ref{cones}), 
surrounding the real axis. Being an entire 
function of infinite order, however, it must grow doubly 
exponenentially in some other directions, such as the 
even-numbered sectors (\ref{sects}) for $\zeta(z)$ as in 
(\ref{wform}). This raises the issue of the relation 
between Minkowski and Euclidean formulations. 

We may follow the standard path of Euclidean field 
theory: the theory is defined through the Euclidean 
path integral for the action (\ref{act}), correlation 
functions are computed, and finally continued to 
Minkowski space by analytic continuation in the 
external momenta. 

For ordinary local (gauge) field theories ($h = 0$), this is 
sometimes purely formally justified as 'Wick rotation of the 
integration contour' of the functional integral. But, in fact, 
actual rigorous justification at the non-perturbative level  
has only been obtained for 'simple' theories.\footnote{These 
are the 'reconstruction theorems' showing that Minkowski 
correlation functions obtained by analytic continuation 
will obey the Wightman axioms if the theory obeys the 
usual Euclidean axioms such as reflection positivity. 
There are, of course, no such 
rigorous results for $4$-dim gauge theories.} Within the 
perturbative loop expansion, however, the procedure is 
indeed justified on a graph by graph basis. It may at first 
appear that this is no longer the case when we allow 
$h \neq 0$. But this is not so.  Analytic 
continuation in the external momenta of the result of 
computation with Euclidean Feynman rules again formally 
agrees with the result of using Minkowski Feynman rules 
and Wick-rotating integration contours according to the 
following procedure.

For convenience, we choose the gauge-fixing weight function
$w(z) = \oh(z)$, and adopt the Feynman gauge $\xi = 1$. 
The basic point is that the vertices (\ref{V}) and the 
factor $\oh^{-1}$ in the propagator (\ref{prop}) 
do not contain any singularities. In the computation of some 
arbitrary graph then we proceed in a standard fashion 
introducing Schwinger 
parameters for the scalar propagator $(k^2 + i\epsilon)^{-1}$ 
in each propagator (\ref{prop}). This allows one to 
successively perform the momentum integrals. The integral over 
internal momentum $k$ is of the form 
\beq
\int d^dk\: {\cal V}\left( \{p^\mu\},\{l^\mu\},\{k^\mu,k^\nu,
\cdots,k^\lambda\}\right) \: \exp\:  [ \: A(x)k^2 + B(p,l,x)
\cdot k \: ] \quad\:.\label{kint}
\eeq
Here $A$ depends only on the Schwinger parameters $x$, while 
$B$ depends on $x$ and also {\it linearly} on the other loop 
momenta $l^\mu$, and external momenta $p^\mu$. ${\cal V}$ stands 
for the product of all vertex and $\bar{h}^{-1}$ factors, and 
thus consists of sums of products of polynomials and entire 
functions. We may then reexpress (\ref{kint}) as 
\beq 
{\cal V}\left( \{p^\mu\},\{l^\mu\},\{\frac{\partial}{\partial 
B^\mu},\frac{\partial}{\partial B^\nu},\cdots,
\frac{\partial}{\partial B^\lambda}\}\right)\: \int d^dk\: 
\exp\: [ \:A(x) k^2 + B(p,l,x)\cdot k\: ] \quad \:.\label{kint1}
\eeq
The Gaussian integral can now be performed by translation 
$k^\mu \to k^\mu - B/A$\ , scaling $k^\mu \to k^\mu/A^{-1/2}$, 
and finally Wick rotation to obtain: 
\beq
i(\frac{\pi}{A})^{d/2} {\cal V}\left( \{p^\mu\},\{l^\mu\},
\{\frac{\partial}{\partial B^\mu},\frac{\partial}{
\partial B^\nu},\cdots,\frac{\partial}{\partial B^\lambda}\}
\right)\: \exp\:( - B^2/A)\quad\:.\label{kint2}
\eeq
The momenta again appear only quadratically and linearly in 
the exponent. Performing the differentiations and picking 
another loop momentum $l_\mu = k^\prime_{\mu}$, we have 
the $k^\prime$-integration in the form (\ref{kint}). Continuing 
in this way all momenta integrations are then 
performed.\footnote{Some regularization, e.g. dimensional, or 
cutting an $\epsilon>0$ off the Schwinger parameter 
integration region, is, of course, implicitly used so 
that this series of steps be always  
well defined. Since in the UV regions all vertices behave as 
polynomials, any of a number of conventional schemes may be 
used.}

We stress that the above is nothing but the standard use of 
Schwinger parameters. It goes through in the present context 
because the usual polynomial vertices are generalized to  
entire functions which again do not introduce any singularities 
in the integrands. The steps (\ref{kint}) - (\ref{kint2}) may 
in fact be viewed here as a definition of the computational 
rules relating Minkowski to Euclidean Feynman rules.

\section{Unitarity and Causality}
\setcounter{equation}{0}

As discussed in section 4, the tree-level propagator 
is arranged to have, at the chosen renormalization scale, 
no gauge-invariant unphysical poles. We now turn 
to the issue of unitarity and causality to any 
order in the loop expansion. Again, with the convenient 
choice $w(z)=\oh(z)$ for the gauge weight function, 
the propagator (\ref{prop}) in configuration space is: 
\begin{equation}
{\bf D}_{\mu\nu ab}(x - x^{\prime}) = -\delta_{ab}\: 
\left( g_{\mu\nu} - (1 - \xi)\: 
\frac{\partial_{\mu}\partial_{\nu}}{\Box} \right)\: 
\oh^{\;-1}(\Box)\: D(x - x^\prime) \label{cop}
\end{equation}
where $D(x)$ is the usual bare massless scalar propagator.  
$D(x)$ has the decomposition 
\begin{equation}
D(x-x^\prime) = \theta(x^0-x^{0\;\prime}) D^+ (x-x^\prime) + 
\theta(x^{0\;\prime}- x^0) D^-(x-x^\prime) \ \, \label{sp} 
\end{equation} 
where 
\begin{equation}
D^{\pm}(x) = \frac{1}{(2\pi)^4}\int d^4k \: e^{-ikx} \: 
D^{\pm}(k) = \frac{1}{(2\pi)^4}\int d^4k \: e^{-ikx} \: 
2\pi\theta(\pm k^0) \delta(k^2)   
\label{enf}
\end{equation} 
are the usual $\pm$ energy functions. (\ref{sp}) implies that 
$D(x)$ obeys the KL representation.

Substituting (\ref{sp}) into (\ref{cop}) gives for 
$x^0 \neq x^{0\;\prime}$\ : 
\begin{eqnarray} 
{\bf D}_{\mu\nu ab}(x-x^\prime) & = & \theta(x^0-x^{0\;\prime})
\oh^{\;-1}(\Box)D^{+}_{\mu\nu ab}(x-x^\prime,\xi) \nonumber\\
 &  & \; \; + \; \theta(x^{0\;\prime} - x^0)\oh^{\;-1} (\Box) 
D^{-}_{\mu\nu ab}(x-x^\prime,\xi) \;\;, \;\;
   \;\;\;\;\;\;\;\;\;\;x^0 \neq x^{0\;\prime} \label{pdec}
\end{eqnarray}
with 
\begin{equation}
D^{\pm}_{\mu\nu ab}(x, \xi) \equiv  -\delta_{ab}\: 
\big( g_{\mu\nu} - (1 - \xi)\: 
\frac{\partial_{\mu}\partial_{\nu}}{\Box} \big)\:D^{\pm}(x) 
\; \: . \label{venf}
\end{equation}

${\bf D}_{\mu\nu ab}(x-x^\prime)$ no longer 
satisfies this decomposition at $x^0 = x^{0\;\prime}$ where 
the r.h.s. of (\ref{pdec}) differs from ${\bf D}_{\mu\nu ab}$  
by $x^0$-contact terms, i.e. terms proportional to 
$\delta(x^0-x^{0\;\prime})$, and its derivatives, resulting 
from the action of the derivatives in (\ref{cop}) on the 
$\theta$-functions in (\ref{sp}). Now the derivation of 
results such as unitarity and causality conditions via 
largest time equations rely on the decomposition (\ref{pdec}). 
Equal times regions may then, in some cases, require special 
consideration as the 
contact terms present a technical, but 
for the most part innocuous, complication.\footnote{The 
precise treatment of such contact terms in field 
theory is, in general, regularization dependent. Actually 
(\ref{sp}), and hence 
(\ref{cop}) with $D(x-x^\prime)$ given as (\ref{sp}), are 
strictly derived only for $x^0\neq x^{0\prime}$. 
This reflects, in the operator formalism, 
the arbitrariness in the definition of the $T$-product, which is 
completely specified only when its arguments are diffferent. The 
standard convention is that the l.h.s. and r.h.s. of 
(\ref{sp}), (\ref{cop}) also coincide 
in an infinitesimal neighborhood of equal times, 
with the l.h.s. defined independently as the appropriate 
Green's function. Alternative conventions amount 
to the addition of local counterterms in 
the Lagrangian. For detailed discussion see \cite{BS}.} 

Recall first the case of ordinary gauge theories, i.e. 
take $\oh = 1$. \ We write 
$D_{\mu\nu ab}$ for (\ref{cop}) with $\oh=1$. 
When the theory possesses gauge invariance, 
explicit consideration of equal time contact terms becomes 
unnecessary since the contribution of such terms must cancel 
in any physical amplitude, as indicated by the presence of 
gauges where they are absent. In particular, in the Feynman 
gauge, $\xi=1$, all derivatives are eliminated, and 
one indeed has 
\begin{equation}
D_{\mu\nu ab}(x) = \theta(x^0) D^+_{\mu\nu ab}(x) 
+ \theta(-x^0)D^-_{\mu\nu ab}(x)  \label{Fpdec}
\end{equation}
where $D^{\pm}_{\mu\nu ab}(x) \equiv 
D^{\pm}_{\mu\nu ab}(x, \xi=1)$, 
valid for all $x^0$. For the ghost propagator $D_{ab}
(x) \equiv \delta_{ab}D(x)$, the corresponding equation 
follows trivially from (\ref{sp}) with $D_{ab}^\pm(x) 
\equiv \delta_{ab}D^\pm(x)$. Given (\ref{Fpdec}), one 
may proceed to derive \cite{V},\cite{tHV} the largest 
time equation, and hence cutting rules leading 
to unitarity conditions for physical amplitudes; 
and then note that, by virtue 
of the WI, these equations continue to hold if one 
replaces $D^{\pm}_{\mu\nu ab}$ with (\ref{venf}) 
for arbitrary $\xi$.

To follow the same procedure in the case we are 
considering here, when $\oh \neq 1$, 
appears at first somewhat problematic. Since the action 
of $\oh(\Box)$ induces contact terms in all parts of the 
propagator, these cannot be cancelled by gauge invariance 
alone. (This, of course, reflects the fact that $h \neq 0$ 
is an actual, gauge invariant modification of the usual 
gauge theory action.)

It is, however, not difficult to circumvent this problem. The 
trick is to use $D_{\mu\nu ab}(x)$ as the bare propagator, and 
include the $\oh(\Box)$ factors in the vertices where the 
propagator line ends. More precisely, write (\ref{prop}) as
\begin{equation}
{\bf D}_{\mu\nu ab}(k) = \oh^{\;-1/2}(k^2) \: 
D_{\mu\nu ab}(k) \: \oh^{\;-1/2}(k^2) \ \ \ \ , \label{psplit}
\end{equation}
and define vertices:
\begin{eqnarray}
\hat{V}_{\mu_1a_1\ldots\mu_n a_n}(k_1, \ldots, k_n) 
& \equiv & V_{\mu_1a_1\ldots\mu_n a_n}(k_1, \ldots, k_n) \:
\prod_{i=1}^n \; \oh^{\; -1/2}(k^2_i) \nonumber\\
\hat{V}_{\mu abc}(k,p,q) & \equiv & V_{\mu abc}(k, p, q) \: 
\oh^{\;-1/2}(k^2) \nonumber\\
\hat{J}_{\mu a}(k) & \equiv & J_{\mu a}(k) \:
\oh^{\;-1/2}(k^2) \qquad \ , \label{vsplit}
\end{eqnarray} 
Here $ V_{\mu_1a_1\ldots\mu_n a_n}$ are the perturbative 
vertices from the expansion of the action (1) (Section 3), 
with $V_{\mu abc}(k,p,q)$ the ghost-ghost-gauge boson vertex, 
and we redefined external sources $J$ by inserting 
$\oh^{\;-1/2}$ factors. For this  
to work, it is, of course, crucial that $\oh(z) \neq 0$ 
and singularity-free for all $|z| < \infty$ ; 
we set $\oh^{\;-1/2} \equiv \exp (-1/2H)$. This 
assignment of $\oh$ factors to vertices is, of course 
not unique, but a convenient, symmetric choice.

Consider now diagrams constructed from propagators $D$, and 
vertices $\hat{V},\; \hat{J}$, the FP propagator remaining 
unchanged. We will refer to these rules as the 'alternative' 
rules. It is immediately seen that 
for any diagram contributing to any $n$-point 
function between arbitrary sources the same result is 
obtained with these rules as with the 
original rules ( propagator ${\bf D}$, vertices $V,\; J$): 
\begin{equation}
G_{\alpha_1\ldots\alpha_n}(p_1,\ldots,p_n)\;
\prod_{i=1}^{n}\; J_{\alpha_i}^i(p_i) = 
\hat{G}_{\alpha_1\ldots
\alpha_n}(p_1,\ldots,p_n)\;\prod_{i=1}^{n}\;
\hat{J}_{\alpha_i}^i(p_i) \ \ \ \ \ . \label{gfeq}
\end{equation}
(Here the label $\alpha_i$ stands for all polarization, 
Lorentz and group indices pertaining to the $i$-th 
leg; and different sources $J^i$ may be chosen for 
each leg.)

For $S$-matrix elements all legs, in addition to be 
truncated, must be put on mass-shell,\footnote{As usual 
with massless poles, to avoid on-shell IR divergences, 
the mass-shell condition is taken with an infinitesimal 
mass, denoted $k^2=0^+$.} and all 
wave-functions appropriately chosen. Now 
\begin{equation}
\oh^{\;1/2}(0) = \oh^{\;-1/2}(0) = 1 \qquad , 
\label{msh}
\end{equation}
so all $\oh$ factors on external on-shell legs 
are actually irrelevant; and 
the residue of the (dressed) gauge boson propagator 
at the pole:  
\begin{eqnarray} 
k^2{\bf \tilde{D}}_{\mu\nu ab}(k)\mid_{k^2=0^+} & = & 
k^2 \oh^{\;-1/2}(k^2)\: \tilde{D}_{\mu\nu ab}(k)\:
\oh^{\;-1/2}(k^2)\mid_{k^2=0^+}\nonumber\\
     & = & k^2\tilde{D}_{\mu\nu ab}\mid_{k^2=0^+} 
\nonumber\\
      & = & \left[ -g_{\mu\nu} - (Z_3 - 1)\:(
g_{\mu\nu ab} - \frac{k_{\mu}k_{\nu}}{k^2})
\right]_{\big| k^2=0^+} \label{res}
\end{eqnarray} 
is the same for both sets of rules. (In (\ref{res}) 
${\bf\tilde{D}}$ and $\tilde{D}$ denote the full 
propagators in the two sets of rules.) 
The residue defines the wave-function 
renormalization constant $Z_3$. Each physical external 
wave-function, in addition to being of physical 
polarization, must be normalized by a factor of 
$1/\sqrt{Z_3}$ in order to correctly account for the 
contribution of self-energy insertions on external 
legs.\footnote{Note that these factors are needed 
to make the $S$-matrix gauge invariant whether or not 
the $Z_3$'s are UV finite.} Truncating and putting all 
legs on shell then, we have the equality for amplitudes:
\begin{eqnarray}
A(p_1,\ldots.p_n)\mid_{p^2_i=0^+} & = & \prod_{i=1}^n
\;J_{\alpha_i}^i(p_i)\;\prod_{i=1}^n\;p_i^2 \; 
G_{\alpha_i\ldots\alpha_n}(p_1,\ldots,p_n)\mid_{p_i^2=
0^+}\nonumber\\
    & = & \prod_{i=1}^n
\;J_{\alpha_i}^i(p_i)\;\prod_{i=1}^n\;p_i^2 \; 
\hat{G}_{\alpha_i\ldots\alpha_n}(p_1,\ldots,p_n)\mid_{p_i^2=
0^+}\qquad. \label{msamp}
\end{eqnarray} 
Note that (\ref{msamp}) holds for arbitrary $J$'s of 
any polarization; and by the equality (\ref{res}), 
it continues to hold for correctly normalized physical 
wave functions. Furthermore, (\ref{gfeq}) - (\ref{msamp}) 
are also valid directly in the renormalized theory, 
since the counterterms in (\ref{actr}) are included in the set 
of vertices $V$. Once more, the absence of singularities 
in $h(k^2)$, and the fact that its functional form 
does not change under renormalization, are crucial for 
obtaining the simple relations (\ref{gfeq}) - (\ref{msamp}). 
These relations allow one to view any diagram as 
constructed according to either set of rules. One may 
then proceed pretty much as in the case of ordinary 
gauge theories.

{\sl Unitarity} \\ 
We construct amplitudes using the alternative 
rules in the Feynman gauge. The bare propagator $D_{\mu\nu ab}$  
then satisfies the decomposition (\ref{Fpdec}). The vertices 
$\hat{V}$, eq. (\ref{vsplit}), are real for real momenta, 
and given by entire functions possessing no singularities, 
in particular no poles or cuts anywhere in the finite 
complex plane. It follows (Appendix C) that the Veltman 
largest time equation \cite{V} holds, which in turn implies 
the cutting equation (generalized Cutkosky rule):\\
\parbox{13.5cm}{\epsfysize=4cm \epsfxsize=12cm 
\epsfbox{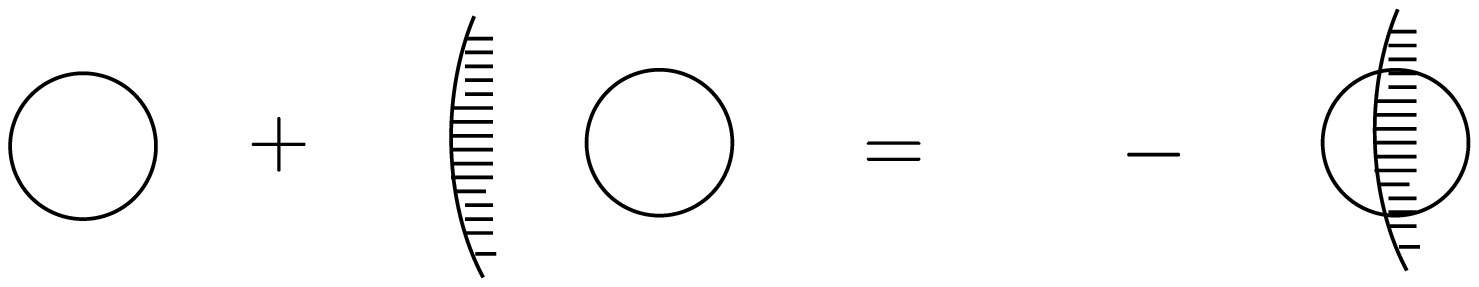}} \hfill 
\parbox{1.5cm}{\begin{equation}\label{cuteq}
\end{equation}}\\
(\ref{cuteq}) is a general cutting rule that applies 
to a single 
diagram, or to any collection of diagrams represented by the 
blob, and for arbitrary external wave-functions and momenta. 
On the shaded side of the 'cuts', each explicit factor of $i$ 
assigned to each vertex and propagator is changed to $-i$, 
and each $i\epsilon$ in each propagator to $-i\epsilon$. 
The cut blob on the right hand side stands for the 
sum over all possible cuts of 
the diagram, or of each diagram in the collection of diagrams 
represented by the blob. A 
possible cut is obtained by cutting propagator lines so that 
the vertices on the shaded (unshaded) side of the cut form a 
connected region containing at least one outgoing (ingoing) 
external line. The rules for cut lines are given by:\\
\parbox{8cm}{\epsfysize=5.5cm \epsfxsize=11cm 
\epsfbox{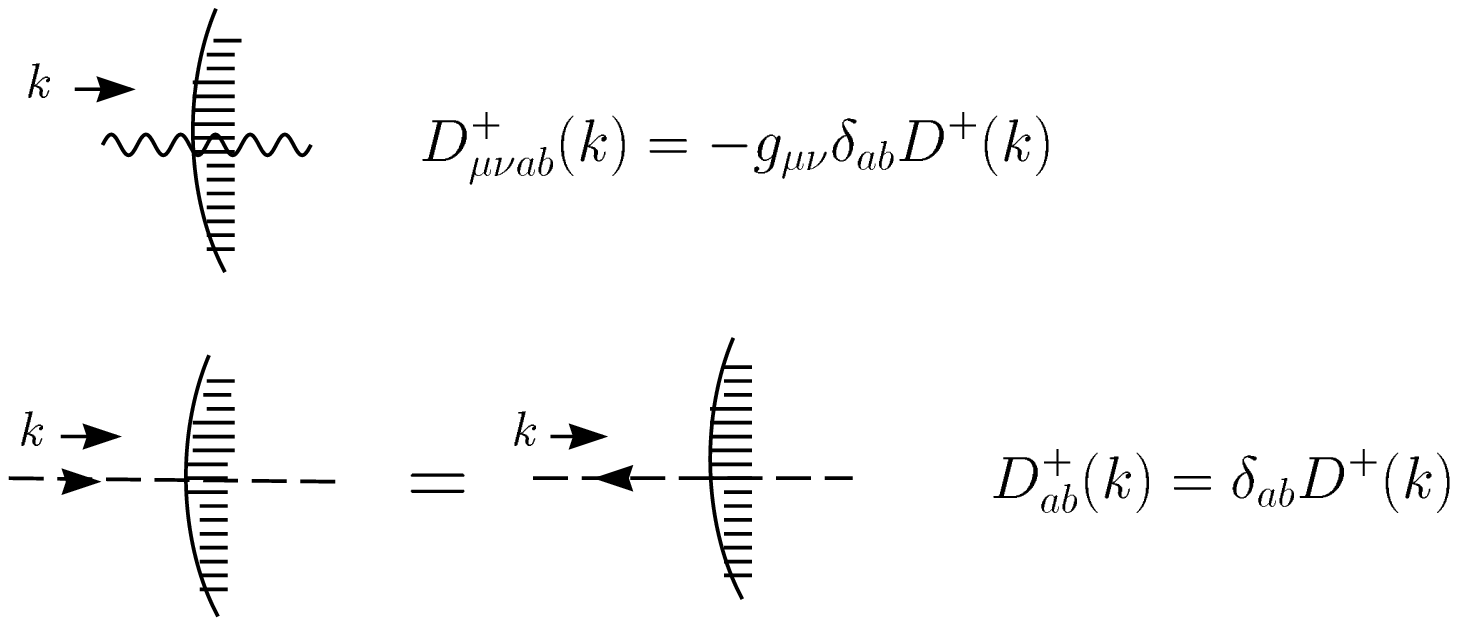}} 
\hfill \parbox{1.5cm}{\begin{equation}\label{cutru}
\end{equation}}\\
\nopagebreak
for gauge field and ghost lines, respectively. Although  
conveniently derived in terms 
of the alternative rules, the result (\ref{cuteq}), once 
obtained, may be trivially viewed 
either in the alternative or the original rules by use 
of (\ref{gfeq}). Note, in particular, that cut legs are 
on shell, so, from (\ref{msh}) 
\[\oh^{\;-1/2}(k^2)\:D^{\pm}_{\mu\nu ab}(k)\:\oh^{\;-1/2}
(k^2)\mid_{k^2=0^+} = D^{\pm}_{\mu\nu ab}(k) \mid_{k^2=0^+}
\qquad \ ,\]
and there is no distinction between a cut ${\bf D}_{\mu\nu ab}
(k)$\  and a cut $D_{\mu\nu ab}(k)$ propagator.

To establish unitarity equations one lets the blobs in 
(\ref{cuteq}) include all diagrams to a certain order 
that contribute to a given process, with all external 
legs truncated and on shell. The physical 
S-matrix is furthermore defined only between external 
physical gauge bosons:\\
\parbox{14cm}{\epsfysize=3cm \epsfxsize=14cm 
\epsfbox{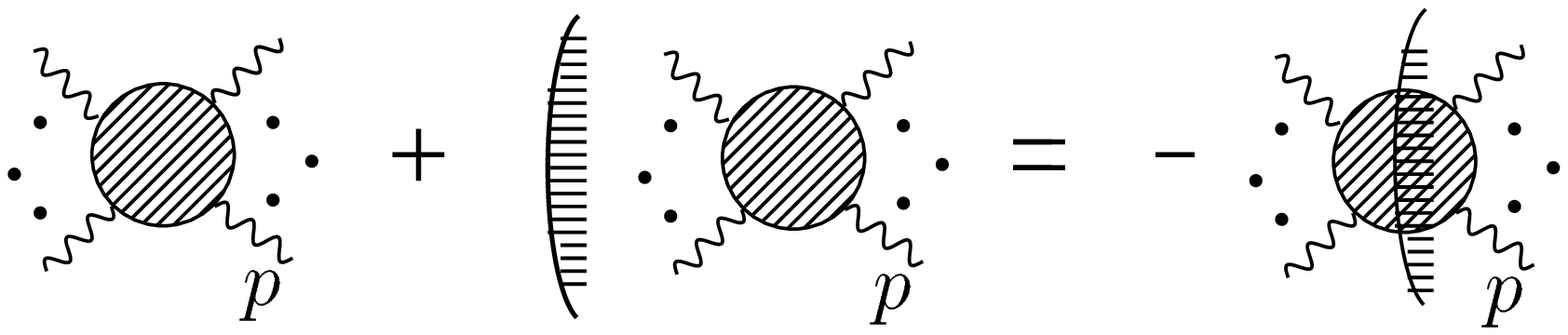}} \hfill 
\parbox{1cm}{\begin{equation}\label{cuteq1}
\end{equation}}\\
The label $P$ denotes physically polarized on-shell gauge 
bosons. Note that since the Lagrangian (\ref{act}) is real 
(hermitean), the rules for diagrams on the shaded side 
in (\ref{cuteq1}) are indeed those for $S^{\dagger}$. In 
the sum over intermediate state cuts on 
the r.h.s. of (\ref{cuteq1}),  
however, the cuts, as given by (\ref{cutru}), include cuts 
over gauge bosons of unphysical 
polarizations, as well as ghosts. Therefore, 
physical unitarity will hold only if these 
unphysical contributions cancel leaving only a sum 
over physical transverse cuts given by:\\
\parbox{5cm}{\epsfysize=1.5cm \epsfxsize=3cm 
\epsfbox{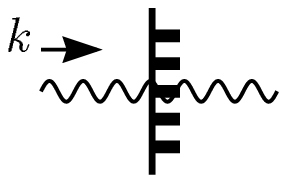}}     
\hfill \parbox{8cm}{\begin{equation}
D_{\mu\nu ab}^{tr\;+}(k) = -g_{\mu\nu}^{tr}\;\delta
_{ab}D^{+}(k) \qquad .\label{trcut}
\end{equation}}\\
$D_{\mu\nu ab}^{tr\;\pm}$ arises from summation 
over only the two physical polarizations satisfying 
$e_{\mu}^Ik^\mu = 0 \; ;\;  e_\mu^I\eta^\mu 
=0 \quad (I=1,2)$, and 
\begin{equation}
\sum_{I=1}^2 \: e^I_\mu e^I_\nu = -g_{\mu\nu}^{tr} 
 \equiv -g_{\mu\nu} - \left[ \frac{1}{2}\:(k_\mu\bar{k}
_\nu + k_\nu\bar{k}_\mu) + k^2\eta_\mu\eta_\nu\right]\;
\frac{1}{(k\cdot\eta)^2 - k^2} \quad,\label{phpol1}
\end{equation}
where $\eta^\mu$ is a timelike unit vector used to fix a 
timelike polarization direction, and 
\beq
\bar{k}_\mu \equiv k_\mu - 2(k\cdot\eta)\eta_\mu \qquad. 
\label{bark}
\eeq

Having 
established the cutting equation (\ref{cuteq1}), 
as well as the Ward identities, eq. (\ref{WI}), however, 
the demonstration of the cancellation of unphysical cuts 
is actually no different than that in the 
case of ordinary local gauge theory. This is clear since 
this demonstration relies only on the gauge, or BRS, 
invariance of the action. It is usually stated at the 
formal level of the path integral independence of the 
gauge-fixing term, which, of course, holds here as well. 
Since, however, we derived the cutting equation by means 
of the split (\ref{psplit})-(\ref{vsplit}) on a graph by 
graph basis, one should, for completeness, check the 
cancellation also diagrammatically. We outline the
derivation in Appendix B, which, starting from the WI 
(\ref{WI}), verifies explicitly (see (\ref{PhUn})) that 
indeed the sum over cuts on the r.h.s. of (\ref{cuteq1}) 
reduces to the sum over only physical cuts (\ref{trcut}).

{\sl Causality} \\ 
Again, working in terms of the 
alternative rules, the validity of the largest time 
equation implies (Appendix C) that the Bogoliubov 
Causality Condition (BCC) \cite{BS}, \cite{V} 
is satisfied:\\[0.5cm]
\parbox{14cm}{\epsfysize=3cm \epsfxsize=14cm 
\epsfbox{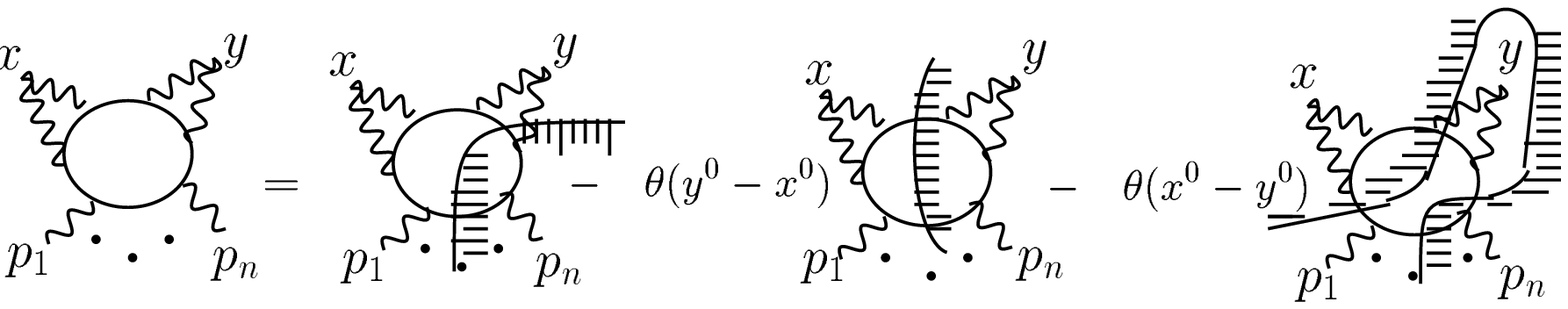}}     
\hfill \parbox{1cm}{\begin{equation}\label{bcc}
\end{equation}}\\[0.5cm]
The blob represents a diagram, or a collection of diagrams, 
contributing to the $n+n_1+n_2$ - point Green's function, 
with $n$ external (truncated) legs, and with $n_1$ 
and $n_2$ legs joined at the space-time 
points $x$ and $y$ by $n_1$ - and $n_2$ - point 
vertices $V$, respectively.\footnote{(\ref{bcc}) is 
actually shown with $n_1=n_2=2$.} $n_1=1$ and/or $n_2=1$ is the 
case of external source vertices at $x$ and/or $y$. 
The cut blobs stand for the sum over all cuts with the 
positions of the two vertices at $x$ and $y$ as shown. 
(Again, though the equation is conveniently derived 
in the alternative rules, it is equally well viewed in 
terms of the original rules by (\ref{gfeq}) and 
(\ref{msamp}).) 

The physical meaning of (\ref{bcc}) is as follows. The first 
term on the r.h.s. is a (set of) cut diagram(s) 
representing a contribution to the product $SS^{\dagger}$, 
with the vertices at $x$ and $y$ both in the 
diagram(s) making up the $S$ 
factor of the product. We may now apply equation (\ref{bcc}) 
to this diagram(s) for the $S$ factor. Iterating this 
procedure, the r.h.s. of (\ref{bcc}) can be reduced 
entirely to a sum of two groups of terms: one 
group multiplied by $\theta(y^0 - x^0)$ and 
containing only cuts forcing positive energy 
flow from $x$ to $y$, the other group 
involving the opposite combination.   

The vertices at $x$ and $y$ act as (multileg) external 
sources and the legs emanating from them correspond in 
general to particles off-shell. This is actually what 
gives the obvious intuitive meaning to the above 
physical interpretation of future directed positive flow 
since, of course, space-time points cannot be precisely 
pinpointed by wave-packets representing particles 
near mass-shell. 

Integrating (\ref{bcc}) over $x$ and $y$ converts 
(\ref{bcc}) to an equation (now entirely in momentum 
space) for a (set of) diagram(s) contributing to an 
$n$ - point amplitude.\footnote{The two vertices at 
$x$ and $y$ are now internal vertices.  
They may also be taken to be external vertices: if 
they are originally chosen as $n_1+n^\prime_1$-  and 
$n_2+n^\prime_2$-point vertices, respectively, then 
multiplication by the appropriate external wave-functions 
and integration over $x$ and $y$ converts (\ref{bcc}) 
to an equation for an $n+n^\prime_1+n^\prime_2$ - point 
amplitude.} The l.h.s. is precisely the (set of) 
diagram(s) for the amplitude in question and is  
expressed by the r.h.s. in terms of cut graphs in what is 
in fact a dispersion relation in non-covariant 
form.\footnote{It is noteworthy that such a dispersion 
relation can be written for any individual diagram. In 
some cases this relation may be converted to a more 
conventional dispersion relation in some external Lorentz 
invariant as the dispersed variable. For scalar theories 
this is developed in \cite{R}.} In the presence of derivative 
interactions, however, some care must be exercised in 
converting (\ref{bcc}) into 
a dispersion relation by integration over $x$ and $y$. 
This is because (\ref{bcc}) was strictly derived for 
$x^0 \neq y^0$. The action of derivatives at $x, y$ 
on the $\theta$-function factors, resulting into 
$\delta$-functions and $\delta$-function derivatives 
which give a finite measure contribution to the equal 
times integration region, must then be properly taken into 
account as explained in Appendix C. The important special 
case of the $2$-point function is considered below.

\section{The 2-point function} 
\setcounter{equation}{0}
Consider (\ref{bcc}) for $n_1=n_2=1,\: n=0$, i.e. for 
the two point function betwen sources at $x,\; y$. 
To lowest approximation, where the blob is a single 
bare propagator line joining $x$ and $y$, (\ref{bcc}) 
is nothing but eq. (\ref{pdec}), sandwiched between 
arbitrary sources, in Feynman gauge ($\xi=1$).
Consider next (\ref{bcc}) for the $2$-point function 
between sources at $x$ and $y$ including an arbitrary 
number of insertions of the self-energy 
\begin{equation}
\Pi_{\mu\nu ab}(x - x^\prime) = i\delta_{ab}\;\int\; 
d^4k\; e^{-ik(x-x^\prime)}(k_\mu k_\nu - g_{\mu\nu}k^2)
\pi(k^2)\qquad .\label{self}
\end{equation}
Summing over all insertions between sources of physical 
polarization, so as to eliminate the physically inessential 
longitudinal terms at the outset, one arrives at the 
BCC equation:\\[0.3cm]
\parbox{14cm}{\epsfysize=3cm \epsfxsize=14cm 
\epsfbox{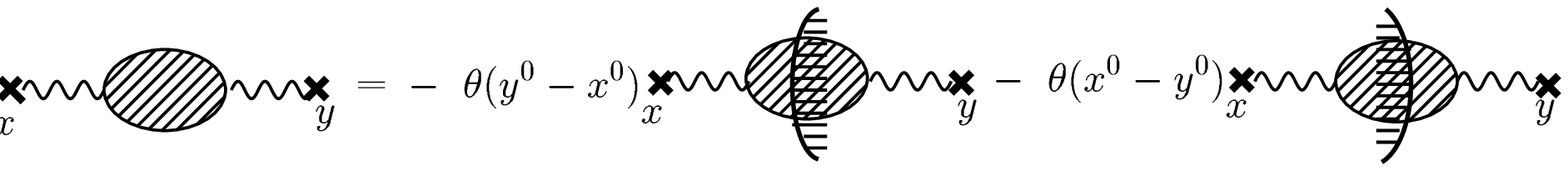}}     
\hfill \parbox{1cm}{\begin{equation}\label{2bcc}
\end{equation}}\\[0.3cm]
In (\ref{2bcc}) the crosses indicate the sourses, and the 
shaded blob stands for the physical transverse 
full propagator given by 
\begin{eqnarray}
{\bf \tilde{D}}^{tr}_{\mu\nu ab}(x - y) & = & \delta_{ab}
\;\sum_{I=1,2} \:\int \;\frac{d^4k}{(2\pi)^4} \; e^I_\mu(k) 
e^I_\nu(k)\;\frac{1}{\oh(k^2)}\;\frac{i}{k^2+i\epsilon}
\;\frac{\T e^{-ik(x-y)}}{\left[1+\frac{\T\pi(k^2)}
{\T(2\pi)^4}\oh^{\;-1}(k^2)\right]} \nonumber\\
    & \equiv & \delta_{ab}\;\sum_{I=1,2}\;e_\mu^I
(i\partial_x)e_\nu^I(-i\partial_y){\bf\tilde{D}}(x-y)
\nonumber\\
   & = & \delta_{ab}\left( -g_{\mu\nu} - 
\frac{\partial_{x\mu}\partial_{x\nu}}{
{\mbox{\boldmath$\partial$}}^2}
\right)(1-\delta_\mu^0)(1-\delta_\nu^0){\bf\tilde{D}}(x-y)
\label{fullp}
\end{eqnarray}
(the last equality valid in the frame where $\eta^\mu=
(1,0,0,0)$, no sum over $\mu,\nu$). Explicitly, 
\begin{eqnarray}
iJ^{\mu a}(i\stackrel{\to}{\partial_x}){\bf\tilde{D}}^{tr}
_{\mu\nu ab}(x-y)iJ^{\nu b}(-i\stackrel{\gets}{\partial_y})
& = & -iJ^{\mu a}(i\stackrel{\to}{\partial_x})
\left[\theta(x^0-y^0){\bf\tilde{D}}^{+}
_{\mu\nu ab}(x-y)\right. \nonumber\\ 
 &  & + \left.\theta(y^0-x^0){\bf\tilde{D}}^{-}
_{\mu\nu ab}(x-y)\right](-i)J^{\nu b}
(-i\stackrel{\gets}{\partial_y})\ \ ,\nonumber\\
   & &  \qquad \qquad \qquad x^0\neq y^0 \label{fpbcc}
\end{eqnarray}
where 
\bea
{\bf\tilde{D}}_{\mu\nu ab}^{\pm} & = & \delta_{ab}
\;\sum_{I=1,2}\;\int\;\frac{d^4k}{(2\pi)^4}\;
e^I_\mu(k)e^I_\nu(k)\oh^{\;-1}(k^2)\theta(\pm k^0)
\rho(k^2,k^2)e^{-ik(x-y)} \nonumber\\
   & \equiv & \delta_{ab}\;\sum_{I=1,2}\;e^I(i\partial_x)
e^I(-i\partial_y) {\bf\tilde{D}}^\pm (x-y) \label{fenf}
\eea
with
\bea
\rho(k^2,q^2) & = & 2\pi Z_3\delta(k^2) + \theta(k^2)
\sigma(k^2,q^2) \label{speca}\\
Z_3 & = & \left[1 + \frac{\pi(0)}{(2\pi)^4}\right]^{-1}
\label{specb}\\
\sigma(k^2,q^2) & \equiv & \frac{1}{(2\pi)^4}
\frac{2{\rm Im}\,\pi(k^2)}{\left| 1 + 
\frac{\T\pi(k^2)}{\T(2\pi)^4}
\oh^{\;-1}(q^2)\right|^2}\frac{1}{\T k^2\:\oh(q^2)}
\label{specc} 
\eea
(\ref{fenf}) is the result of cutting the propagator 
(\ref{fullp}):\\[0.3cm]
\parbox{14cm}{\epsfysize=3cm \epsfxsize=14cm 
\epsfbox{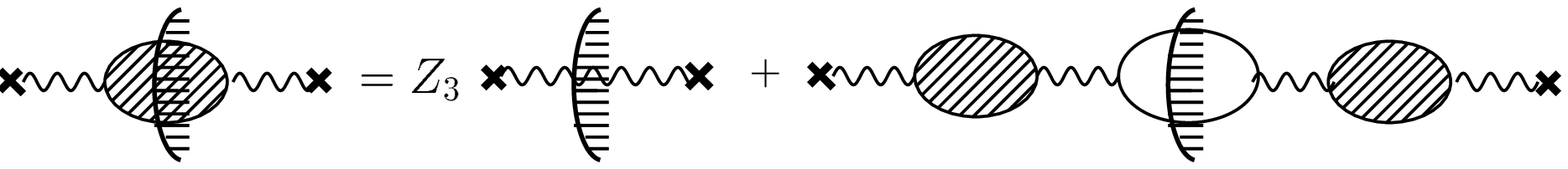}}     
\hfill \parbox{1cm}{\begin{equation}\label{cutp}
\end{equation}}\\[0.3cm]
The first term on the r.h.s. of (\ref{cutp}) represents 
the $\delta$-function term in $\rho(k^2,k^2)$, and is 
the contribution to the cut of the pole in (\ref{fullp}). 
The second term on the r.h.s. of (\ref{cutp}) represents 
the substitution of the second term in (\ref{speca}) into 
(\ref{fenf}). It is easily seen that this gives precisely 
the structure depicted graphically in the second term in 
(\ref{cutp}) with the cut self-energy given by:\\
\parbox{3.5cm}{\epsfysize=2cm \epsfxsize=3cm 
\epsfbox{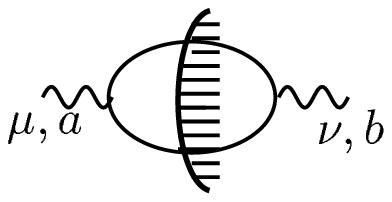}} \hfill
\parbox{11.5cm}{\begin{equation}
\Pi^\pm_{\mu\nu ab}(k) = \delta_{ab}\left(-g_{\mu\nu}k^2 
+ k_\mu k_\nu\right)\theta(k^2)\theta(\pm k^0)2{\rm Im}\,
\pi(k^2) \label{selfcut}
\end{equation}}\\

Noting that $\rho(k^2,q^2)$ is non-vanishing only for 
$k^2\geq0$, one can write: 
\bea
{\bf\tilde{D}}^\pm(x-y) & = & \frac{1}{2\pi}\oh^{\;-1}
(-\Box_x/\Lambda^2)\int_0^\infty d\kappa \rho(\kappa,
-\Box_x)\,\int \frac{d^4k}{(2\pi)^4}\,e^{-ik(x-y)}2\pi
\theta(\pm k^0)\delta(k^2-\kappa) \nonumber\\
   & = & \frac{1}{2\pi}\oh^{\;-1}(-\Box_x/\Lambda^2)\;
\int_0^\infty\;d\kappa\;\rho(\kappa, -\Box_x)\Delta^\pm
(x-y,\kappa) \label{fenfa}
\eea
where $\Delta^\pm(x-y,\kappa)$ is the $\pm$-energy function 
of the free scalar field of mass $m^2=\kappa$. Substituting 
in (\ref{fpbcc}) one obtains 
\bea
{\bf\tilde{D}}(x-y) & = & \frac{1}{2\pi}\oh^{\;-1}
(-\Box_x/\Lambda^2)\;\int_0^\infty\,d\kappa\;\rho(\kappa,
-\Box_x)\nonumber\\
     & & \quad\qquad\cdot\left[\theta(x^0-y^0)
\Delta^+(x-y,\kappa) + 
\theta(y^0-x^0)\Delta^-(x-y,\kappa)\right]\nonumber\\
   & = & \frac{1}{2\pi}\oh^{\;-1}
(-\Box_x/\Lambda^2)\;\int_0^\infty\,d\kappa\;\rho(\kappa,
-\Box_x) \Delta(x-y,\kappa) \qquad, \label{invKL}
\eea
where $\Delta(x-y,m^2)$ is the free scalar propagator. Again, 
this derivation from (\ref{2bcc})-(\ref{fpbcc}) strictly 
follows for $x^0\neq y^0$, but the result {\it as written in} 
(\ref{invKL}) correctly includes also the point $x^0=y^0$ 
(Appendix C). (\ref{invKL}) is the (generalized) KL
representation for the invariant function ${\bf\tilde{D}}
(x-y)$ in the decomposition (\ref{fullp}). Reverting to 
momentum space, one obtains the dispersion relation 
for the physical transverse full propagator: 
\begin{eqnarray}
{\bf\tilde{D}}^{tr}_{ijab}(p) & = & \frac{i}{(2\pi)^4}
\delta_{ab}\left(\delta_{ij} - \frac{p_ip_j}{|{\bf p}|^2}
\right)\oh^{\;-1}(p^2/\Lambda^2)\;\frac{1}{2\pi}
\int_0^\infty\;d\kappa\; \frac{\rho(\kappa,p^2)}{p^2 - 
\kappa + i\epsilon} \nonumber\\
   & = & \frac{i}{(2\pi)^4}
\delta_{ab}\left(\delta_{ij} - \frac{p_ip_j}{|{\bf p}|^2}
\right)\;{\bf\tilde{D}}(p) \qquad.\label{pdis}
\end{eqnarray}

The BCC-equation (\ref{2bcc}) for (\ref{fullp}) was obtained 
by summing the BCC-equation for the $2$-point function with 
$n$ self-energy insertions over all $n$. This summation 
is exhibited by expanding the denominators in (\ref{specb}), 
(\ref{specc}) to write:
\begin{equation}
Z_3=\sum_{n=0}^\infty \;Z_3^{(n)}\quad, \qquad 
\sigma(k^2,p^2)=\sum_{n=0}^\infty\;\sigma^{(n)}(k^2)\;
\oh^{\;-n}(\frac{p^2}{\Lambda^2}) \label{specex}
\end{equation} 
with $Z_3^{(0)}\equiv 1\;,\: \sigma^{(0)}(k^2)\equiv 0\;,$ 
so that 
\begin{equation}
{\bf\tilde{D}}(p) = \oh^{\;-1}(\frac{p^2}{\Lambda^2})\:
\sum_{n=0}^\infty \;\left\{\frac{Z_3^{(n)}}{p^2+i\epsilon} 
\;+\; \frac{1}{2\pi}\oh^{\;-n}(\frac{p^2}{\Lambda^2})\;
\int_0^\infty\,d\kappa\;\frac{\sigma^{(n)}(\kappa)}
{p^2 -\kappa +i\epsilon}\right\}\:.\label{pdisex}
\end{equation} 

Applying the BCC-equation (\ref{bcc}) also to the self-
energy (\ref{self}) itself, and proceeding as above, 
one derives the standard dispersion relation: 
\bea
\Pi_{\mu\nu ab}^{sub}(k) & = & i\delta_{ab}(k_\mu k_\nu 
- g_{\mu\nu}k^2)\left(\pi(k^2) - \pi(0)\right)\nonumber\\
   & = & i\delta_{ab}(k_\mu k_\nu - g_{\mu\nu}k^2)
\frac{1}{2\pi}\;\int_0^\infty\;d\kappa\;
\frac{k^2\sigma_0(\kappa)}{(k^2-\kappa+i\epsilon)(\kappa 
-i\epsilon)} \quad,\label{selfdis}
\eea
where we defined $\sigma_0(\kappa)\equiv2{\rm Im}\,
\pi(\kappa)$. The dispersion integral over 
$\sigma_0(\kappa)$ does not converge, hence requiring 
the subtracted dispersion (\ref{selfdis}), i.e. 
renormalization, the subtraction of 
$\pi(0)$ being one particular convenient choice of 
counterterm (giving unit residue to the renormalized 
dressed propagator (\ref{fullp})). 
Note that the self-energy subtraction does not affect 
the imaginary part of $\pi(k^2)$.\footnote{This is, of 
course, the basic point of renormalizability by local 
counterterms.}

(\ref{2bcc}), (\ref{fpbcc}) hold with the self-energy 
(\ref{self}) computed in any approximation. 
When an appropriate (gauge invariant) set of graphs 
is included, the imaginary part $\sigma_0(k^2)$, and 
hence the spectral density $\rho(k^2,p^2)$ 
in (\ref{pdis}) is positive. Indeed, from the 
demonstration of physical unitarity above, only 
physical positive residue particles propagate then in 
the cut (\ref{selfcut}). Eqs. (\ref{invKL}),  
(\ref{pdis}), (\ref{pdisex}), (\ref{selfdis}) now 
make explicit the fact that this positivity is 
indeed consistent with the improved 
asymptotic behavior of the propagator in this 
superrenormalizable theory.

Recall that in the case of the bare propagator,  
represented by the $n=0$ term in 
(\ref{pdisex}), the improved asymptotic behavior 
is due to the $\oh(k^2/\Lambda^2)$ factor, which 
does not contribute to the absorptive part - the 
only contribution is the usual $\delta$-function 
due to the pole at $k^2=0$. (\ref{pdis}), (\ref{pdisex}) 
show that this continues to hold when radiative 
corrections are included: the external momentum 
dependence in the spectral density $\rho(\kappa,p)$ 
is {\it entirely} due to the $\oh^{\;-1}$ factors 
in the {\it bare propagators} connecting the 
self-energy insertions; and {\it the improved 
asymptotic behavior is entirely due to these 
$\oh^{\;-1}$ factors.} The self-energy (\ref{self}) 
does, of course, depend implicitly on the function 
$h$ through its occurance in the propagators and 
vertices inside the self-energy loops forming 
$\Pi_{\mu\nu ab}$; but (\ref{self}) itself does {\it not} 
contribute to the improved convergence of (\ref{fullp}): 
according to our power counting above (section 3), 
$\pi(k^2)$ has constant or at most logarithmic 
dependence on large $p$. (Note that this is the 
usual asymptotic behavior of self-energy diagrams 
in ordinary gauge theories). This is why the dispersion 
integral (\ref{selfdis}) requires precisely 
one subtraction. Were $\Pi_{\mu\nu ab}(p)$ to have a 
higher than $p^2$ asymptotic dependence, thus 
contributing to the improved convergence of (\ref{fullp}), 
additional subtractions would be required corresponding 
to the addition of counterterms containing polynomials 
in derivatives of degree larger than two. These, as 
we saw above, would introduce additional poles other 
than $k^2=0$ in (\ref{prop}), some of necessarily negative 
residue. Equivalently, the functional form of 
$\oh(x)$ would change under renormalization, the 
renormalized $\oh(x)$ acquiring zeroes. Thus 
the renormalizability and unitarity properties of 
the theory are closely connected.\footnote{The fact 
that the leading asymptotic behavior of the dressed 
propagator is entirely due to its bare propagator part, 
i.e. self-energy insertions do not contribute to it,  
appears to be a general feature of superrenormalizable 
theories - cp. $\phi^3$-theory in $4$ dimensions.}

In summary, eqs. (\ref{pdis}) - (\ref{selfdis}) show 
explicitly that  the relation 
\begin{equation}
\lim_{p^2\to\infty} p^2 {\bf\tilde{D}}^{tr}(p) = 0 
\label{asym}
\end{equation}
is indeed consistent with the result that $Z_3 \geq 0$, 
\ $\sigma_0 > 0$, 
i.e. that only positive residue physical particles 
propagate in the intermediate state cuts.

\section{Gravity}
\setcounter{equation}{0}

In analogy to (\ref{act}), we consider the gravitational 
action: 
\bea
\cL& =& \sqrt{-g}\:\left\{\:\frac{\beta}{\kappa^2}R 
- \beta_2(R_{\mu\nu}R^{\mu\nu} - \frac{1}{3}R^2) + \beta_0R^2 + 
\lambda \right.\nonumber\\
 & &+\left.\left(R_{\mu\nu}\,h_2(-\frac{\nabla^2}{\Lambda^2})
\,R^{\mu\nu} -\frac{1}{3}R\,h_2(-\frac{\nabla^2}{\Lambda^2})\,R
\right) - R\,h_0(-\frac{\nabla^2}{\Lambda^2})\,R
\:\right\}\nonumber\\
 & & -\frac{1}{2\xi}f^\mu[g]w(-\frac{\Box}{\Lambda^2})
f_\mu[g] + \bar{c}^\mu M_{\mu\nu}c^\nu \label{gact}
\eea
where $\nabla^2=\nabla^\mu\nabla_\mu$ and $\Box$ denote the 
covariant and ordinary D' Alembertian, respectively, and 
$f_\mu[g]$ is the gauge-fixing function with gauge-term weight 
$w$. In (\ref{gact}) two terms introducing the transcendental 
entire functions $h_2$ and $h_0$ have been added to the general 
4-th order renormalizable gravitational Langrangian. Expanding, 
for convenience, about flat space: 
\beq
\sqrt{-g}\,g^{\mu\nu} = \eta^{\mu\nu} + \kappa \phi^{\mu\nu}
\qquad,
\eeq
and taking $f^\mu = \partial_\nu \phi^{\nu\mu}$, 
the bare propagator is of the form:
\bea 
D_{\mu\nu\kappa\lambda}(k)& =& \frac{i}{(2\pi)^4}\frac{1}{k^2+
i\epsilon}\left( \frac{2P^{(2)}_{\mu\nu\kappa\lambda}(k)}{
\beta - \beta_2\kappa^2k^2 + \kappa^2k^2h_2(k^2/\Lambda^2)}
\right. \nonumber\\
 & &  \qquad \qquad \qquad + \left.\frac{4P^{
(0)}_{\mu\nu\kappa\lambda}(k)}{-\beta + 6\beta_0\kappa^2k^2 - 
6\kappa^2k^2h_0(k^2/\Lambda^2)}\right)\nonumber\\
  &  & \qquad\quad+ (\xi-\mbox{proportional trace and 
longitudinal parts})\;,\label{gprop}
\eea
where $P^{(2)}$ and $P^{(0)}$ denote spin-2 and spin-0 
projections, respectively. 

The structure of the vertices involving $h_2$ or $h_0$ is 
completely analogous to (\ref{V}) - (\ref{v}). Again, in 
obvious matrix notation, we have:
\bea
V^{(N)}(x; x_1,\ldots,x_N)   & = & {\rm tr}\left(
\left.\frac{\delta^{n^{\prime\prime}}{\bf 
R}[g]}{\delta\bphi^{n^{\prime\prime}}}
\right|_{\bphi=0}\cdot v^{(n)}(x,\partial_x; x_1,\ldots,x_n)
\cdot\left.\frac{\delta^{n^{\prime}}{\bf R}[g]}{\delta\bphi^{
n^{\prime}}}\right|_{\bphi=0}\right) 
\nonumber\\
 & \equiv & {\rm tr}\;{\bf R}^{(n^{\prime\prime})}\cdot v^{(n)}
\cdot{\bf R}^{(n^\prime)}\quad, \qquad N=n^{\prime}+n+
n^{\prime\prime} \label{gV}
\eea 
where 
\bea
v^{(n)} & = & \frac{\delta^n}{\delta\bphi^n}\: 
\sum_{r=0}^\infty \; a_r\; \left(-\frac{\nabla^2[g]}{
\Lambda^2}\right)^r_{|\bphi=0} \nonumber\\
& = & \sum_{l=1}^n \sum_\sigma\; \sum_{r=l}^{\infty}\: a_r \; 
{\cal S}_{r,l,n}^\sigma 
\left( \left(-\frac{\Box}{\Lambda^2}\right)^{(r-l)}, \left(
\frac{1}{\Lambda^2}\frac{\delta^b}{\delta\bphi^b}[-\nabla^2 
+ \Box]_{|\bphi=0}\right)^l
\right) \nonumber\\
  & \equiv &  \sum_{l=1}^n \sum_\sigma \:v^\sigma_{l,n}(x,
\partial_x;\; x_1, \ldots,x_n) \quad . \label{gv} 
\eea  
In (\ref{gv}), $\{a_r\}$ are the coefficients in the expansion 
(\ref{hexa}) of either $h_2$ or $h_0$. ${\cal S}^\sigma_{r,l,n}$ 
stands for the sum of all 
possible ways of distributing $(r-l)$ powers of $(-\bBox)$ in 
the $l+1$ positions among an ordered sequence, indexed by  
$\sigma$, of $l$ factors of $\frac{\delta^b}{\delta\bphi^b}
[-\nabla^2 +\bBox]|_{\bphi=0}$. Note that $[-
\nabla^2 +\Box]$ can generate any number of $\bphi$ legs, 
so now $b \in [1,n-l+1]$, and the number of ordered sequences 
labeled by $\sigma$ is 
\beq
\Gamma = l^{(n-l)}\;l! \quad . \label{goseq}
\eeq 
This structure $v_{l,n}^\sigma$ is again given by (\ref{struc}), 
and the analysis in Appendix A applied to (\ref{gv}) gives 
its behavior for arbitrary leg momenta configuration. In 
particular, if any subset of the leg momenta is of order 
$k$ with $k$ growing arbitrarily large, the leading asymptotic 
behavior of $v_{l,n}^\sigma$ is given by a sum of terms that 
grow at most either as 
\beq
 h_m^{(s)}(k^2)\;k^{2s} \: \:\:,\qquad 
0\leq s \leq l \quad,\quad m=0,\;2\quad, \label{gas1}\\
\eeq
\beq
\mbox{or}\qquad \qquad\: \frac{1}{k^2}k^{2n^{{\rm int}}} \qquad
 \qquad\qquad \qquad\qquad\qquad. \label{gas2}   
\eeq 
We impose the requirement that both $h_2$ and $h_0$ exhibit 
the same, at most polynomial asymptotic behavior in a 
neighborhood of the real axis, so that 
\beq
\gamma \equiv \lim_{|z|\to\infty \atop Im\: z \to 0} \left( 
\frac{\ln |h_2(z)|}{\ln|z|}\right) = 
\lim_{|z|\to\infty \atop Im\: z \to 0} \left( 
\frac{\ln |h_0(z)|}{\ln|z|}\right) \label{ggamma}
\eeq
exists. Power counting (Appendix A) then shows that, 
provided $\gamma \geq 2$, UV divergences arise solely due 
to (\ref{gas1}), and the superficial degree of divergence 
of any 1PI graph without external ghost legs is bounded by 
\beq 
\delta_G \leq 4 - 2\gamma (L-1) \quad.  \label{gdivdeg}
\eeq 
Thus, if $\gamma \geq 3$, only 1-loop diagrams are 
superficially divergent. Graphs 
with any external ghost lines are superficially convergent for 
all $L$. Thus only gauge-invariant 1-loop divergences occur -  
the theory is superrenormalizable and is renormalized by the 
addition of all gauge-invariant 4-th order 1-loop counterterms:
\beq
\cL_R = \cL + \sqrt{-g}\:\left\{\:\frac{\beta(Z-1)}{\kappa^2}R 
- \beta_2(Z_2-1)(R_{\mu\nu}R^{\mu\nu} - \frac{1}{3}R^2) + 
\beta_0(Z_0 - 1)R^2 + \lambda(Z_\lambda-1) \right\} 
\;, \label{gactr}
\eeq
where all the couplings in (\ref{gactr}) now signify 
renormalized couplings at some scale $\mu$. Again, let us note 
that $h_2$ and $h_0$ are not altered by renormalization. 

We now define 
\beq
\oh_2(z) \equiv \beta - \beta_2\kappa^2\Lambda^2z+ 
\kappa^2\Lambda^2zh_2(z) \quad, \quad \oh_0(z) \equiv \beta - 
6\beta_0\kappa^2\Lambda^2z + 6\kappa^2\Lambda^2zh_0(z)
\eeq
An explicit form of the functions $h_2$, $h_0$ follows,  
as before, by imposing conditions (i) - (iii) of section 4 on 
each of them. We take 
\beq
h_2(z) = \frac{1}{\kappa^2\Lambda^2}\;\left(\:\frac{\alpha 
\,(e^{H(z)} - 1) + \alpha_2 z}{z}\:\right) \quad, \qquad 
h_0(z) = \frac{1}{\kappa^2\Lambda^2}\;\left(\:\frac{\alpha 
\,(e^{H(z)} - 1) + \alpha_0 z}{6z}\:\right)\quad.\label{ghform}
\eeq 
Here $H(z)$ is again given by (\ref{Hform}) but with 
the replacement $\gamma \longrightarrow (\gamma +1)$. 
(\ref{wform}), (\ref{Hform1}) again provide an explicit 
realization. $\alpha,\;\alpha_2,\;\alpha_0$ are parameters. 
Assume that the theory has been 
renormalized at some scale $\mu_0$. Then the bare propagator 
(\ref{gprop}) will possess no gauge-invariant pole other 
than the transerse massless physical graviton pole if we set  
\beq 
\alpha = \beta(\mu_0) \quad , \qquad 
\alpha_2\;\frac{1}{\kappa^2\Lambda^2} = \beta_2(\mu_0) \quad , 
\qquad  \alpha_0\;\frac{1}{\kappa^2\Lambda^2} = 6\beta_0(\mu_0) 
\quad .\label{alfix}
\eeq
This may be viewed as fixing the scale \footnote{Cp. 
the discussion in the vector gauge theory case above.}
$\Lambda^2$ in terms of the (Planck) scale $1/\kappa^2$ 
present in the theory: 
\beq
\Lambda^2 = \left(\frac{\alpha_2}{\beta_2(\mu_0)}\right)\;
\frac{1}{\kappa^2}\qquad,\label{Lfix}
\eeq 
and fixing 
\beq
\beta(\mu_0) = \alpha \quad , \qquad 
\frac{\beta_2(\mu_0)}{\alpha_2} = 
\frac{6\beta_0(\mu_0)}{\alpha_0}\qquad .\label{alfix1}
\eeq
Note that any split between $\beta_2$, resp. $\beta_0$, and 
a constant part in the entire function $h_2$, resp. $h_0$ 
(i.e. a coefficient $a_0$ in the expansion (\ref{hexa})) 
is actually a renormalization convention.\footnote{As already 
pointed out, the crucial point is that there are no 
counterterms that renormalize $\{\:a_n\:|\:n\geq 1\:\}$ 
altering the non-trivial dependence of $h_2$ and $h_0$ on 
their argument.} Thus, again, there is no loss in 
generality in assuming that we work with a renormalization 
prescription such that (\ref{alfix1}) holds at 
some conveniently chosen scale $\mu_0$. If now $\mu_0$ is 
taken as the renormalization point, $\oh_2(z) = \oh_0(z) = 
\beta(\mu_0)\exp\:H(z)$, and only the physical massless 
spin-2 pole occurs in the bare propagator. If another 
renormalization scale $\mu$ is chosen, the previous 
discussion (section 4) applies: (\ref{gprop}) will 
acquire poles which, however, will cancel in the dressed 
physical propagator since, by RG invariance, the shift in 
the bare part leading to their appearance will cancel 
against a corresponding shift in the self-energy.   

The discussion of unitarity and causality, 
and the 2-point function of the previous two sections goes 
through here as well in a closely analogous manner.

\section{Discussion}
\setcounter{equation}{0}

We have constructed a class of vector gauge and 
gravitational theories by including in the action a 
series of derivative terms representing the expansion of 
a transcendental entire function. The original Lagrangian to 
which these terms are added is renormalizable. The new 
terms then render it superrenormalizable, provided the 
entire function(s) determining the new vertices possess 
appropriate asymptotic behavior. Additional constraints 
follow from the requirement  that no (gauge- and 
RG-invariant) unphysical poles are introduced in the 
propagators. These conditions then fix the class of 
allowed entire functions. Cutting 
equations may then be derived within the loop expansion,     
which allow one to verify that no unphysical cuts occur 
in the intermediate states to any order. The unitarity 
and superrenormalizability properties are closely related as 
discussed in section 7: the self-energy 
insertions are irrelevant to the improved UV behavior of 
the propagator. 
  
The arbitrariness within the class of allowed entire 
functions is that of a theory admitting a class of  
potentials. This is nothing unusual in field 
theory.\footnote{The superrenormalizable $P(\phi)_2$ models 
and supergravity potentials are familiar 
examples.} The interesting point here is that 
this potential depends on derivatives of the fields. 
The study in this paper was within the perturbative loop 
expansion. Harder to investigate issues outside perturbation 
theory, such as global stability, may lead to further 
constraints on the allowed class of functions, and/or the 
inclusion of additional terms in the action. 

It is straightforward to introduce minimally coupled 
non-self-interacting matter, i.e. fermions in gauge theories,  
and any gravitating matter with action bilinear in the 
matter fields in the gravitational case. It is immediately 
seen that again only 1-loop graphs can be superficially 
divergent. In particular, any matter loop embelished by 
internal gauge boson or graviton propagators is superficially 
convergent; and so is any graph with any external matter 
field legs. The superrenormalizability properties of the 
theory are thus not altered by the matter 
coupling.\footnote{The abelian version of (\ref{act}) 
minimally coupled to fermions is a particularly 
interesting case. Since there are 
no photon self-interaction vertices, the only modification 
introduced by the $h$-term in (\ref{act}) is in the photon 
propagator. This is then a version of QED with smooth UV 
behavior and apparently no UV Landau pole.} 
Note in this connection that anomalies still 
occur as usual, since the relevant 1-loop graphs, 
such as fermion triangle graphs, will be divergent.   
 
Self-interacting matter fields, such as scalars, do 
introduce multi-loop divergences through subgraphs 
of multiloops of these matter lines. 
Thus the theory is rendered merely renormalizable. Whether 
this introduces any inconsistency remains to be investigated. 
In this connection one may, of course, consider modifying 
also the matter Langrangian in a manner analogous to 
(\ref{act}). For actions only bilinear in the matter fields, 
such a modification actually introduces nothing new since 
it may be absorbed in a redefinition of the matter fields. 
Only interaction terms trilinear or higher in matter fields, 
if present, will be affected. In this manner one may 
perhaps obtain superrenormalizable theories also 
in the presence of Higgs scalars. In fact, the troublesome 
quadratic scalar divergences might be completely 
eliminated. This question will be addressed elsewhere. 

Perhaps more interesting is the fact that the theories 
introduced here can be expected to have a spectrum of bound 
states, thus creating their own 'matter': these may be bound 
states of gauge bosons or gravitons, as well as of externally 
introduced matter. Indeed, at length scales of order 
$1/\Lambda$ or smaller, the inverse bare gauge boson or 
graviton propagator is given by a polynomial of 
degree $2+2\gamma$ or $4+2\gamma$, 
respectively, corresponding to a tree-level confining or 
ultraconfining potential between particles. This is, of 
course, reflected in the smooth UV behavior of the 
theory. At scales larger than $1/\Lambda$, (\ref{act}) 
and (\ref{gact}) reduce to the usual theories, and revert 
to $1/k^2$ tree-level propagators corresponding to 
Coulomb potentials. The exact shape of the potential 
barrier between the two regimes is determined by the 
particular choice of the function $H$ in (\ref{Hform}). 
For sufficiently large $\Lambda$, and exponential 
fall-offs as in (\ref{wform}), 
it can be made extremely steep. Systems bound by the short 
distance confining potential will thus be effectively 
permanently confined, and forming string-like excitations 
with almost linearly rising spectrum. Note that the formation 
mechanism for such states is perturbative, indeed classical, 
since, as we saw, the short distance behavior is completely 
dominated by the bare propagator. It is, therefore, not to be 
confused with any long-distance nonperturbative confining 
interaction arising from the nonperturbative IR dynamics 
in nonabelian gauge theories (\ref{act}). 

Such tightly bound states, e.g. scalars as gauge boson bound 
states, may have some interesting applications. In the 
gravitational case, the very smooth UV behavior of 
power-like approach to vanishing short-distance interaction, 
and the related nature of the excitation spectrum, have 
some obvious relevance for the problems of the avoidance 
of space-time singularities and of entropy in gravitational 
collapse. 

It is plausible that appropriate supersymmetrization of 
the actions (\ref{act}) and (\ref{gact}) may lead to 
cancellation of the remaining 1-loop divergences, and 
thus perturbatively finite theories.

\section*{Acknowledgements}

Part of this work was done while the author was at the 
Yukawa Institute of Theoretical Physics. The hospitality 
and support of the Institute is gratefully acknowledged. 
I would also like to thank the theory group at the Physics 
Department, University of Tokyo (Komaba), for their 
hospitality and discussions, and Z. Bern and H. Sonoda for 
discussions.
\setcounter{section}{0}  
\renewcommand{\thesection}{\Alph{section}}
\renewcommand{\theequation}
{\mbox{\thesection.\arabic{equation}}}

\section{Appendix - Vertices and power counting}
The general structure of each term in the sum ${\cal S}
^\alpha_{r,l,n}\cdot{\bf F}^{(n^\prime)}$ appearing in 
(\ref{V})-(\ref{v}) is of the form\footnote{To avoid 
notational clutter, in (\ref{struc})-(\ref{Bform}) and 
the rest of this section division of all d'Alembertians 
by $\Lambda^2$ is left implied.}: 
\beq
\sum_{\{n_i\}}\:\;(-\bBox)^{n_1}\;B_1\;(-\bBox)^{n_2}
\;B_2\;(-\bBox)^{n_3}\;\cdots\;(-\bBox)^{n_l}\;B_l\;
(-\bBox)^{n_{l+1}}
\cdot{\bf F}^{(n^\prime)} \quad, \label{struc}
\eeq
where 
\beq
B_i \equiv \frac{\delta}{\delta{\bf A}^{b_i}}\; [\;-{\bf D}^
2 + \bBox\;]_{|{\bf A}=0}\:\quad,\qquad \sum_{i=1}^l\:b_i = 
n\quad,\quad b_i \in \{1,2\}\:.\label{Bform}
\eeq
 
Note that each $B_i$ is the source of either one derivative-
coupled vertex leg contributing a linear power of momentum 
($b_i=1$), or two legs and no powers of momentum ($b_i=2$). 
Let $\delta^{(l)}$ denote the number of derivative-coupled 
(single-legged) $B_i$' s in (\ref{struc}). The following 
relation then holds: 
\beq
\delta^{(l)} + n = 2l \label{derno}
\eeq
The $B_i$'s in (\ref{struc}) are in a fixed 
ordered sequence labeled $\alpha$; the number of such 
sequences is given by (\ref{oseq}). 

Given the sequence of the $l$ $B_i$'s, there are $l+1$ 
positions for distributing $(r-l)$ factors $\bBox$ between 
them. The sums in (\ref{struc}) then are over all sets 
$\{n_i\}$ such that 
\beq
n_1 + n_2 + \cdots +n_{l+1} = (r - l)\:\quad .\label{sets}
\eeq
The number of solutions to (\ref{sets}) is equal to\footnote{
This is the number of colourings of $(r-l)$ indistinguishable 
balls ( $m_j$ balls in the case of (\ref{msets}) below), with 
$(l+1)$ colours ($l_j$ colours in (\ref{msets})), and 
repetitions of any colour allowed.}
\beq
{(l+1) + (r-l) -1 \choose r-l} = {r \choose l}\quad.
\label{slno}
\eeq

Going over to momentum space, let $q_i$ denote the sum of 
momenta of the legs emanating from $B_i$, $i=1, \ldots, l$, 
and $q_{l+1}$ the sum of momenta of the legs from ${\bf F}^{
(n^\prime)}$. Then $(-\bBox)^{n_i}$ generates a 
factor of $(Q_i^2)^{n_i}$, where $Q_i\equiv \left(\sum_{
i^\prime=i}^{l+1}\;q_{i^\prime}\right)\;$ is the sum of the 
vertex momenta incoming through the ${\bf F}^{(n^\prime)}$ 
factor, and all $B_{i^\prime}$'s, $i^\prime = i, \ldots, 
l\;,$ to the right\footnote{Note that 'left' and 'right' can 
be interchanged by overall vertex momentum conservation.} of 
$(-\bBox)^{n_i}$. (We set $Q_{l+1}=q_{l+1}$.) In general, 
for some vertex momenta configurations, equalities 
among some of the linear combinations 
$Q_i$ may occur because of cancellations 
in partial momenta sums. (To examine the convergence 
properties of graphs we will need, according to the proof of 
the power counting theorem, ascertain the behavior of the 
vertices along every hyperplane in the space of the momenta.) 
Let the $(l+1)$ possible positions of the $\bBox$'s in 
(\ref{struc}) be split into $J$ disjoint sets, where the $j$-th 
subset ($j\in [1,J]$) consists of $l_j$ positions all giving 
the same momentum factor $Q_j^2$. Let $n_i^{(j)},\: i=1,
\ldots,l_j$ denote the exponents in (\ref{struc}) in these 
$l_j$ positions, and 
\beq
\sum_{i=1}^{l_j} \: n_i^{(j)} = m_j \:\quad .\label{msets}
\eeq
Then 
\beq
\sum_{j=1}^J \:l_j = l+1\quad , \quad\quad \sum_{j=1}^J \: 
m_j = r - l\quad\:\:,\label{subsets} 
\eeq
where $1\leq J\leq l+1$. So for $J=1$, $l_1=l+1$, $m_1 = 
r-l$ and all $l+1$ positions give the same $Q$; for 
$J=l+1$, each of the $l+1$ positions results in a different 
partial momentum sum $Q_i$, and $l_j=1$, $j=1,\ldots,l+1$. 
There are ${m_j + (l_j-1) \choose m_j}$ solutions to 
the constraint (\ref{msets}). 

The vertex $v^\alpha_{l,n}$ in momentum space is then of 
the form: 
\bea
 & &\left[\sum_{r=l}^\infty \:\;\sum_{\{m_i\}} \: \delta_{r-l,
\;m_1+\cdots+m_J} \; a_r\:\prod_{j=1}^J\;{m_j + (l_j-1) 
\choose m_j}\: Q_j^{2m_j} \right]\: \phi^\alpha_{l,n}(\{Q\})
\nonumber\\
 & & \equiv S_{l,n}\;\phi^\alpha_{l,n} (\{Q\})\quad.
\label{vform}
\eea
Here $\phi^\alpha_{l,n}(\{Q\})$ denotes the product of the 
$l$ factors $B_i$ in (\ref{struc}), and carries all spacetime 
and group indices, here suppressed, on $v^\alpha_{l,n}$. Each 
$B_i$ is a linear combination of $Q_i$ and $Q_{i+1}$ if one-
legged, and has no momentum dependence if two-legged. Now 
\bea
S_{l,n}& = &\prod_{j=1}^J\;((l_j-1)!)^{-1} \frac{d^{(l_j-1)}}{
dQ_j^{2(l_j-1)}}\:\left(\sum_{r-l=0}^\infty\;\sum_{\{m_j\}}\: 
\delta_{r-l,\:m_1+\cdots+m_J} \:a_r\;\prod_{i=1}^J\;Q_i^{2(
m_i+(l_i-1))} \right)\nonumber\\
 &=& \prod_{j=1}^J\;((l_j-1)!)^{-1} \frac{d^{(l_j-1)}}{
dQ_j^{2(l_j-1)}}\:\left(\sum_{\{m_i=0\}}^\infty\:a_{J-1+\sum_i^J
(m_i+l_i-1)} \prod_{i=1}^J\:Q_i^{2(m_i+l_i-1)}\right)
\nonumber\\
 &=& \prod_{j=1}^J\;((l_j-1)!)^{-1} \frac{d^{(l_j-1)}}{
dQ_j^{2(l_j-1)}}\:\left(\sum_{\{m_i=0\}}^\infty \:a_{J-1+\sum_i^J 
m_i} \prod_{i=1}^J\:Q_i^{2m_i}\right)\nonumber\\
 &\equiv& \prod_{j=1}^J\;((l_j-1)!)^{-1} \frac{d^{(l_j-1)}}{
dQ_j^{2(l_j-1)}}\:S_J\;(\;Q_1, \ldots, Q_J\;)\quad\:.
\label{Ssum} 
\eea
To evaluate $S_J$ we may assume that the $Q_i$ are ordered, 
if necessary by relabeling, into descending order: $Q_1^2 > 
Q_2^2 > \ldots > Q_J^2$. Note that $Q_j^2\neq Q_{j^\prime}^2$ by 
assumption. Then
\bea
 & &S_J\;(\;Q_1, Q_2, Q_3,\ldots, Q_J\;) \nonumber\\
 & &= \sum_{\{m_i=0\;|\;i\geq 3\}}^\infty \:\sum_{m_2=0}^
\infty\:\sum_{m_1=m_2}^\infty\:a_{J-1+m_1+m_3+\cdots+m_J} \: 
Q_1^{2m_1} \left(\frac{Q_2^2}{Q_1^2}\right)^{m_2}\;
\prod_{i=3}^J\:Q_i^{2m_i}\nonumber\\ 
 & &= \sum_{\{m_i=0\;|\;i\geq 3\}}^\infty \:\sum_{m_1=0}^
\infty\:\sum_{m_2=0}^{m_1}\: a_{J-1+m_1+m_3+\cdots+m_J} \: 
\left(\frac{Q_2^2}{Q_1^2}\right)^{m_2}\;Q_1^{2m_1} \;
\prod_{i=3}^J\:Q_i^{2m_i}\nonumber\\
 & &= \left(1 - \frac{Q_2^2}{Q_1^2}\right)^{-1}\;\bigg[ 
\:S_{J-1}\;(\;Q_1, Q_3, Q_4,\ldots, Q_J\;)\nonumber\\
 & & \qquad\qquad\qquad\qquad - \left(\frac{Q_2^
2}{Q_1^2}\right)\;S_{J-1}\;(\;Q_2, Q_3, Q_4,\ldots,Q_J\;)\:
\bigg]\quad.\label{recur}
\eea
We may now iterate (\ref{recur}) to perform the rest of the 
sums. After $(J-1)$ iterations one obtains:
\beq
S_J\:(Q_1,Q_2,\ldots,Q_J\;) = \sum_{k=1}^J\: C_k(\{\frac{
Q_j}{Q_i}\}_{i<j\leq k})\:\prod_{m>k}^J\left(
1 - \frac{Q_m^2}{Q_k^2}\right)^{-1}\;\left(\frac{Q_k^2}{Q_1^2}
\right)\;\frac{\tilde{h}_J(Q_k^2)}{Q_k^{2(J-1)}}\label{SJform}
\eeq
where
\beq
\tilde{h}_J(Q^2) \equiv \sum_{m=J-1}^\infty \:a_m Q^{2m}
= h(Q^2) - (1-\delta_{J1})\left(\sum_{m=0}^{J-2}\:a_mQ^{2m}
\right)\:\: ,\label{th}
\eeq
and the $C_k(\{\frac{Q_j}{Q_i}\}_{i<j\leq k})$ ' s are given 
through 
\bea
C_1  &= & 1\label{C1}\\
C_k    &=& - \sum_{s=1}^{k-1} \: C_s\:\prod_{j=s+1}^k\:
\left( 1- \frac{Q_j^2}{Q_s^2}\right)^{-1}\quad,\quad k\geq 2\:.
\label{Cform}
\eea
From (\ref{Cform}) one easily obtains $C_k$, $k\geq 2$, as a 
sum of $2^{k-2}$ terms, each term a product of $k-1$ factors 
and of the form 
$\pm\prod_{\{(ij)\}}\;\left( 1 - Q_j^2/Q_i^2\right)^
{-1} $ over sets of pairs $(ij)$,  with $1 \leq i < j \leq k$; 
for example, for $k=3$, $\{(ij)\} 
= \{(12), (13)\}$ and $ \{(12), (23)\}$. 

Eqs. (\ref{vform}), (\ref{Ssum}), and 
(\ref{SJform})-(\ref{Cform}) allow one to examine the 
behavior of the vertices in any direction in the space of 
the vertex momenta. For  general 
UV power counting, assume that some of the $l+1$ $Q_i$'s 
grow as $Q_i = s_ik+s_i^\prime q \sim s_ik$, where $s_i, s_i^
\prime$ constants, $q$ some finite momentum and $k$ grows 
arbitrarily large. Let then a subset ${\cal K}$ consisting 
of $K$ of the above $J$ sets of equal\footnote{For 
considerations of asymtotic behavior, $Q$ and $Q^\prime$ 
are considered to belong to the same set if 
$Q \sim Q^\prime\sim Q_j \sim c_jk, \; j\in \cK$.} 
$Q$' s grow as $Q_j\sim s_jk$, $j\in {\cal K}$; 
whereas the set ${\cal L}$ of the other 
$L=J-K$ sets stays finite, $Q_j=s_jq$, $j\in {\cal L}$. Let 
\[\sum_{j\in {\cal K}}\:l_j = l^\prime\quad, \quad \sum_{
j\in {\cal L}}\:l_j = l^{\prime\prime}\quad,\qquad l^\prime+l^{
\prime\prime}= l+1\:\:\:,\] 
so $1 \leq K \leq l^\prime\;,\: 0 \leq L \leq l^{\prime\prime}$.
Splitting the summation in (\ref{SJform}) as $S_J = S_K+S_L$, 
one easily obtains from (\ref{SJform})-(\ref{Cform}) the 
leading asymptotic behavior: \pagebreak
\bea
\frac{d^{l^{\prime\prime}-L}}{dq^{2(l^{\prime\prime}-L)}}
\frac{d^{l^\prime-K}}{dk^{2(l^\prime-K)}}\: S_K & \sim & 
\sum_{s=0}^{l^\prime-K} \: \frac{h^{(s)}(k^2)}{k^{2(l-s)}} - 
\frac{a_{J-2}}{k^{2(l+2-J)}}\:\:,\quad J\geq2\label{SKas}\\
 & \sim & h^{(l)}(k^2)\:\:, \qquad\qquad J=K=1\label{SKas0}\\
\frac{d^{l^{\prime\prime}-L}}{dq^{2(l^{\prime\prime}-L)}}
\frac{d^{l^\prime-K}}{dk^{2(l^\prime-K)}}\: S_L & \sim &
y(q^2)\;\frac{1}{k^{2[l^\prime-K+1]}} \label{SLas1}
\eea
where $y(q^2)$ is a combination of $h(q^2)$ and its 
derivatives, and powers of $1/q^2$. 

Since each $B_i$ is at most linear in the momenta, we have 
\beq
\phi^\alpha_{l,n} \leq B\;k^{\tilde{l}^\prime}\quad, \qquad 
\tilde{l}^\prime \leq \delta^{(l)} \leq l\:,\label{Bas}
\eeq
for some constant $B$. If all $B_i$' s are one-legged 
(derivative coupled, $\delta^{(l)}=l$), then for 
non-exceptional vertex momenta configurations $\tilde{l}^
\prime = l^\prime$ for $l+1 \notin \cK$, or $\tilde{l}^
\prime = l^\prime-1$ for $l+1 \in \cK$. In general, 
however, we may have\footnote{Cancellations may occur in the 
partial sums $Q_i$, so that for some $j$, $Q_j \sim q$, but 
$B_j \sim k$. An example would be $q_j = -sk, \: Q_{i-1}
= sk +q$. Similarly, it may be that $Q_j \sim k$, but $B_j
\sim q$. Also, two-legged $B_i$' s contribute 
no powers of momentum. The first case contributes positively, 
the latter two negatively to $\tilde{l}^\prime - l^\prime$.} 
$\tilde{l}^\prime \neq l^\prime, l^\prime-1$.
 
Now, starting at the $i=l+1$ position in (\ref{struc}) count 
the total number of transitions $\cK \to \cK, \: 
\cL \to \cK, \: \cK \to \cL$ encountered as one proceeds to 
$i=1$. (Only transitions between distinct elements count in 
$\cK \to \cK$.) Note that among the first $l$ positions of 
the $Q_i$ in (\ref{struc}) there occur either all $K$ of the 
sets in $\cK$ if $l+1 \notin \cK$,  
or at least $K-1$ of the sets in $\cK$ if $l+1 \in \cK$; 
and only a change $\cK \to \cL$ may contribute positively 
to the difference $\tilde{l}^\prime - l^\prime$. It follows 
that the number of transitions is greater or equal to 
$(K + ( \tilde{l}^\prime - l^\prime))$. Now, for any pair 
$Q_{i-1}, \: Q_i = q_i + Q_{i-1}$, only if $q_i \sim k$, 
i.e if momentum of order $k$ is injected 
into $B_i$, can any of the following 
three occur:   
$\:i$ and $i-1$ belong to distinct elements 
of $\cK$; or $i \in \cK$ while $i-1 \in \cL$;  
or $i \in \cL$ while $i-1 \in \cK$.  
Let then $n^{{\rm int}}$ denote the number of 
legs among the $B_i$ carrying momentum of order $k$. Then 
\beq
n^{{\rm int}} \geq K + (\tilde{l}^\prime - l^\prime) 
\quad\quad,\label{count}
\eeq
and
\beq
\frac{1}{k^{2[l^\prime-K+1]}}\;k^{\tilde{l}^\prime} \leq 
\frac{1}{k^2}\;k^{K+(\tilde{l}^\prime - l^\prime)} \leq 
\frac{1}{k^2}\;k^{n^{{\rm int}}} \quad.\label{ascount}
\eeq
Noting that $0\leq l^\prime-K\leq l$ and 
\[l+2-J = (l^{\prime\prime} - L) + (l^\prime - K + 1)
\geq (l^\prime -K +1) \quad,\]
and combining (\ref{vform}), (\ref{Ssum}), (\ref{SKas})-
(\ref{SLas1}), (\ref{Bas}), (\ref{derno}) and (\ref{ascount}), 
we arrive at eqs. (\ref{as1})-(\ref{as2}) of the main text.

{\it Power Counting}

Let $\delta^F$ denote the number 
of derivative-coupled lines emanating from the ${\bf F}^2$ 
factors in any given vertex in (\ref{act}). Note that 
$0\leq\delta^F\leq2$, and:
\beq
\delta^F + n^F = 4 \label{Fderno}
\eeq
where $n^F, \: 1\leq n^F\leq4,\;$ denotes the number of 
legs emanating from the ${\bf F}$-factors in the vertex. 
Let
\beq
\gamma^{(s)} \equiv \lim_{|z|\to \infty}\lim_{{\rm Im}\: z 
\to 0}\left( \frac{\ln|h^{(s)}(z)\;z^s|}{\ln|z|}\right)
\quad . \label{sgamma}   
\eeq
Consider any 1PI graph $G$ constructed with the vertices 
(\ref{V}). The superficial degree of divergence is:
\bea
\delta_G \leq  \sum_{i\in \cV_1} \:(\delta_i^{F,\;{\rm int}}
+ 2\gamma^{(s_i)}-n_i) + 
\sum_{i\in \cV_2} \:(\delta_i^{F,\;{\rm int}} + n_i^{
{\rm int}} -2)\qquad\qquad & & \nonumber\\
 - \sum_{l^{{\rm int}}}\:(2 + 2\gamma) + 
4(I - V +1) & & \label{supdeg}
\eea 
The index $i$ enumerates the vertices which are split into 
two sets $\cV_1$ and $\cV_2$ with asymptotic behavior 
(\ref{as1}) and (\ref{as2}), respectively. The suffix 
'int'  indicates restriction to internal lines only. $V = 
V_1 + V_2$ is the total number of vertices, and $I$ the 
number of internal lines $\{l^{{\rm int}}\}$.
Noting that $\delta_i^{F,\;{\rm int}} \leq \delta^F_i$, 
and $\delta_i^{F,\;{\rm int}} \leq n_i^{F,\;{\rm int}}$, 
(\ref{supdeg}) may be rewritten as 
\bea 
\delta_G & \leq & \sum_{i\in\cV_1}\: \left( \delta_i^F - n_i 
+ 2\gamma^{(s_i)} - \frac{1}{2}\sum_{l_i^{{\rm int}}}\; 
2(\gamma - 1) - 4\right) \nonumber\\
    & & + \sum_{i\in\cV_2} \: \left( n_i^{F,\;{\rm int}} + 
n_i^{{\rm int}} - \frac{1}{2}\sum_{l_i^{{\rm int}}}\; 
2(\gamma - 1) - 6 \right) + 4 \quad .\label{supdeg1}
\eea 
Using (\ref{Fderno}), and the fact that each vertex must have 
at least two internal legs, and provided $\gamma \geq 1$, 
we obtain 
\bea 
\delta_G & \leq & 
\sum_{i\in\cV_1}\: \left( - N_i + 2(\gamma^{(s_i)} 
-\gamma) + 2\right) + \sum_{i\in\cV_2}\: \left( N_i^{{\rm int}} 
- (\gamma - 1) N_i^{{\rm int}} - 6 \right) + 4 \nonumber\\
 & \leq & \sum_{i\in\cV_1}\: \left(2 - N_i \right) - 
\sum_{i\in\cV_2} \: \left( N_i^{{\rm int}} (\gamma - 2) + 6 
\right) + 4 \label{supdeg3}
\eea 
In the last step we used the important fact that the 
(by construction) polynomial asymptotic behavior of $h(z)$ 
means that 
\beq
\gamma^{(s)} - \gamma \leq 0 \qquad.\label{gamineq}
\eeq 
We set $N_i \equiv n_i^F + n_i$, and similarly for $N_i^{
{\rm int}}$, for the total number of lines and internal 
lines, respectively, out of the $i$-th vertex. Now 
$N_i\geq 2$, so the sum over the $\cV_1$ vertices in 
(\ref{supdeg3}) is always non-positive. In fact we have 
the topological relation
\beq 
\sum_{i\in\cV_1}\: (N_i- 2) = 2( L_1 - 1) + E_1 + I_{12} 
\label{tpl}
\eeq 
where $L_1$ is the number of loops in $G$ constructed 
entirely of vertices in $\cV_1$, and $E_1$ is the number 
of external lines attached to $\cV_1$ vertices, while 
$I_{12}$ is the number of internal lines in $G$ connecting 
a $\cV_1$ and a $\cV_2$ vertex. Thus 
\beq
\delta_G \leq 4 - 6V_2 - \sum_{i\in\cV_2}\:(\gamma-2)N_i^{
{\rm int}}\qquad,
\eeq
and, provided $ \gamma \geq 2$, we always have $\delta
_G < 0$ if $\cV \not= \emptyset$, i.e. $V_2\geq1$ - any graph 
containing one or more $\cV_2$ vertices is superficially 
convergent.

The only superficialy divergent graphs then are those 
containing  solely $\cV_1$ vertices, i.e. the parts in 
(\ref{v}) with asymptotic behavior (\ref{as1}). 
(\ref{supdeg3}) with $\cV_2 = \emptyset$ and 
(\ref{tpl}) with $I_{12}=0$ give:
\beq 
\delta_G \leq 4 - E - 2(L - 1)\qquad \quad.
\label{supdeg4}
\eeq 
A slightly better estimate can be obtained 
directly from(\ref{supdeg}) with $\cV_2=\emptyset$: 
\bea
\delta_G & \leq& \sum_i\: ( \delta_i^F + 2\gamma^{(m_i)} - 
n_i - 4 ) - 2(\gamma-1)I + 4 \nonumber\\
 &\leq& \sum_i\: (-n_i^F-n_i +2 + 2(\gamma^{(m_i)} - \gamma)) 
- 2(\gamma-1)(L-1) + 4\nonumber\\
  & \leq&\sum_i\:(2-N_i) - 2(\gamma -1)(L-1)+ 4\nonumber\\
 &=&4 - 2\gamma(L-1) - E 
\eea 
which is eq. (\ref{divdeg}) in the main text. 

In the same manner, it is straightforward to include in 
(\ref{supdeg}) the usual YM vertices from the 
$\frac{1}{g^2}{\bf F^2}$ and ghost terms in (\ref{act}), 
and with the same result (\ref{divdeg}). In particular, 
graphs with one or more $\frac{1}{g^2}{\bf F^2}$ 
vertices, or any external ghost legs are superficially 
convergent. This is actually rather obvious since 
these vertices and the ghost propagator remain 
unaltered, but the gauge propagator has 
improved UV behavior.

{\it Gravity}

The general structure of each term in the sum $S_{r,l,
n}^\alpha\cdot{\bf R}^{n^\prime}$ is again of the form 
(\ref{struc}) where now
\beq
B_i \equiv \frac{\delta}{\delta\bphi^{b_i}}\:[\:-\nabla^2 + 
\Box\:]_{|\bphi = 0}\quad, \qquad \sum_{i=1}^l\:b_i = n 
\quad,\quad b_i \in [1,(n-l)+1] \:.\label{gBform}
\eeq
Each $B_i$ contributes two derivatives, so $\delta^{(l)}
= 2l$ is the total number of powers of momenta 
contributed by the $l$ $B_i$'s. 

Given a sequence of the $l$ $B_i$'s, the analysis of the 
distribution of the $(-\Box)$'s among the $l+1$ positions 
is as before. Thus starting from (\ref{sets}), 
one repeats the argument leading to the series forms 
(\ref{vform})-(\ref{Ssum}), and the summation result 
(\ref{SJform})-(\ref{Cform}), and asymptotic forms 
(\ref{SKas})-(\ref{SLas1}). We now have 
\beq
\phi^\alpha_{l,n} \leq B\;k^{2\tilde{l}^\prime}\quad, \qquad 
2\tilde{l}^\prime \leq \delta^{(l)} = 2l\:,\label{gBas}
\eeq
while (\ref{count}) remains unchanged. Collecting results as 
before, we obtain eqs. (\ref{gas1})-(\ref{gas2}). 

The superficial degree of divergence of a 1PI graph with 
vertices (\ref{gV}) is now 
\bea
\delta_G \leq  \sum_{i\in \cV_1} \:(\delta_i^{R,\;{\rm int}}
+ 2\gamma^{(s_i)}) + 
\sum_{i\in \cV_2} \:(\delta_i^{R,\;{\rm int}} + 2n_i^{
{\rm int}} -2)\qquad\qquad & & \nonumber\\
 - \sum_{l^{{\rm int}}}\:(4 + 2\gamma) + 
4(I - V +1) & & \label{gsupdeg}
\eea
where $\cV_1$ and $\cV_2$ again denote the sets of vertices 
with asymptotic behavior (\ref{gas1}) and (\ref{gas2}) 
respectively, and $h_2$ and $h_0$ have, by constuction, equal 
index $\gamma^{(s)}$ defined in (\ref{sgamma}). Using 
$\delta_i^{R,\;{\rm int}} \leq 4$, and $\delta_i^{R,\;{\rm 
int}} \leq 2n_i^{{\rm int}}$, proceeding as above (cp eqs. 
(\ref{supdeg})-(\ref{supdeg3})), we now obtain 
\beq 
\delta_G \leq 4 - \sum_{i\in \cV_2}\:( (\gamma -2)N_i^{{\rm 
int}} + 6) \label{gsupdeg1}
\eeq
Hence, if $\cV \not= \emptyset$, we always have $\delta_G < 
0$, provided $\gamma\geq2$. Now, with 
$\cV_2=\emptyset$, (\ref{gsupdeg1}) gives $\delta_G \leq 4$. 
A better estimate, however, follows directly from 
(\ref{gsupdeg}) with $\cV=\emptyset$:
\bea
\delta_G & \leq & \sum_i \:  (\delta_i^{R\;{\rm int}} + 
2\gamma^{(s_i)} - 2\gamma - 4) - 2\gamma (L-1) + 4\nonumber\\ 
  & \leq & 4 - 2\gamma (L-1)
\eea 
which is eq. (\ref{gdivdeg}) in the main text. Again, 
and for the same reason as in the vector gauge theory 
case, the same result is obtained if we include any of 
the original, i.e. $h$-independent, gravitational 
vertices, and ghost vertices.

\section{Appendix - Ward identities}
\setcounter{equation}{0}
The BRS invariance (\ref{BRS}) of the action (\ref{act})
implies, in the standard fashion, the Ward identities (WI): 
\bea
  & &\int\; [D{\bf A}][d\bar{c}][dc]\; \exp\{ i\int\;( \cL + 
{\bf J\cdot A} + \bar{\eta}c + \bar{c}\eta )\}  \nonumber\\
   & &\quad \int\; \left[ J^{a\mu}D^{ab}_\mu c_b + \frac{1}{
\xi}\,w(\frac{-\Box}{\Lambda^2})\, (\partial_\mu A^{a\mu})
\eta^a - \frac{1}{2}\bar{\eta}^af_{abc}c_bc_c\right] = 0 
\label{WI}
\eea
where $J,\:\bar{\eta},\:\eta$ are external sources for the 
gauge and ghost fields.\footnote{It should noted that here 
the absence of gauge dependent divergences means 
that the identities (\ref{WI}) hold directly in the renormalized 
theory in terms of the renormalized Lagrangian (\ref{actr}) 
without further ado - there is no need for a definition 
of a renormalized gauge transformation.} The form of these 
identities is, of course, the same as in ordinary gauge theory 
since they follow only from the gauge (BRS) invariance of 
the action. We can, therefore, be brief. Differentiating 
(\ref{WI}) w.r.t. $\eta(x)$ and then setting $\eta=\bar{\eta}
=0$ gives the ST identities in their usual form, 
from which follows, in particular, the absence of radiative 
corrections to the longitudinal part of the propagator. 
Similarly, the WI for on-shell gauge boson amplitudes, 
needed in the discussion of 
unitarity, also follow by familiar manipulations:\\
\parbox{14cm}{\epsfysize=3cm \epsfxsize=11cm \epsfbox{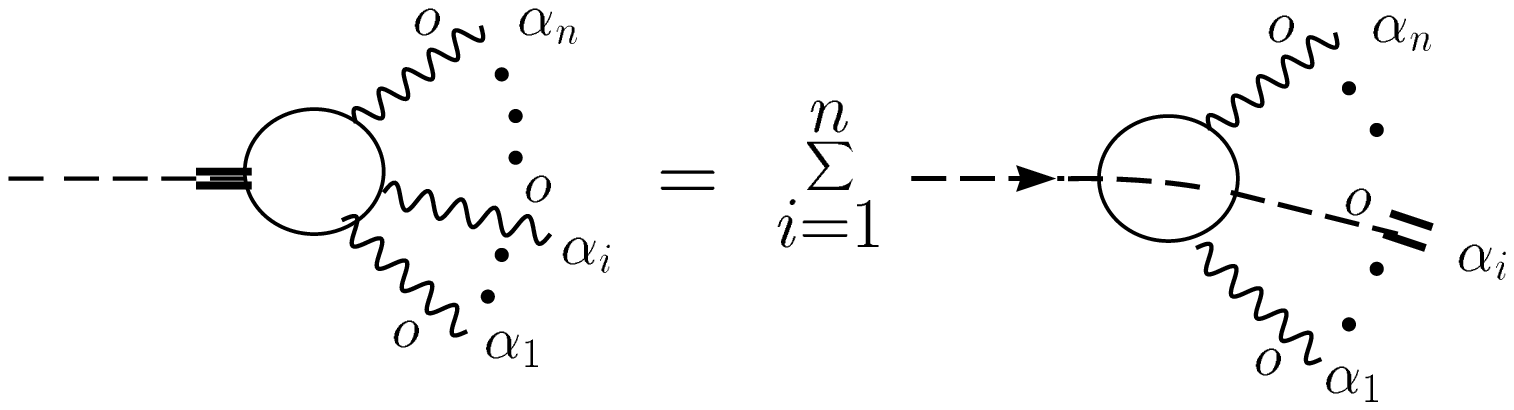}}     
\hfill \parbox{1cm}{\begin{equation}\label{WI1}
\end{equation}}\\ 
The short double line denotes multiplication by $-ip_\mu$ 
where $p_\mu$ is the momentum flowing into the vertex it 
is attached to. The wavy lines indicate truncated external 
gauge boson legs of arbitrary polarization, and the label 
$o$ that they are on-shell. If, say, the j-th leg carries 
physical transverse polarization $e^\mu$, then 
$p_\mu e^\mu=0$, and the corresponding term is absent 
in the r.h.s. sum. 

By further differentiation of (\ref{WI}) w.r.t. $\bar{\eta},
\:\eta$, and repeating the route leading to (\ref{WI1}), 
one obtains analogous 
WI involving any number of external ghost legs. Off-shell, 
such identities look progressively more complicated as the 
number of legs increases. On mass-shell, however, they 
simplify considerably - they essentially reduce to the 
form of (\ref{WI1}), with the blob carrying the additional 
ghost legs and summations over appropriate permutations of 
external lines. 

Having obtained the WI (\ref{WI1}), the cancellation of 
unphysical polarizations and FP ghosts on the r.h.s. 
of (\ref{cuteq1}) can now be explicitly verified.
Introduce, following \cite{tHV}, the auxiliary cuts \\
\parbox{14cm}{\epsfysize=2cm \epsfxsize=13.5cm 
\epsfbox{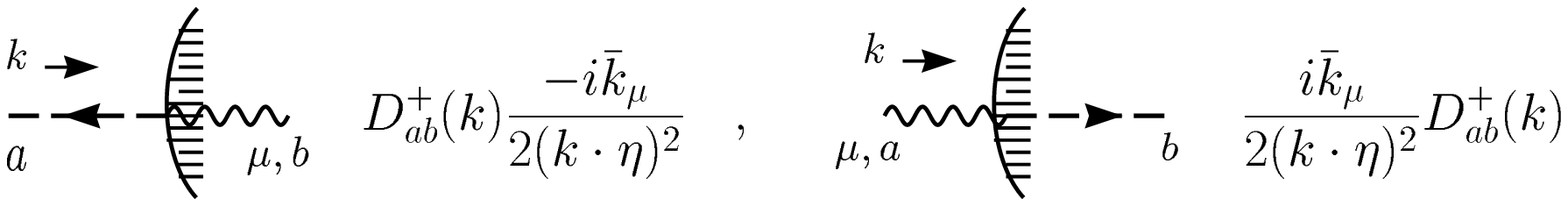}} \hfill 
\parbox{1cm}{\begin{equation}\label{cutaux}
\end{equation}}\\ 
(\ref{phpol1}) on shell, $k^2=0$, then gives the relation\\ 
\parbox{14cm}{\epsfysize=2.5cm \epsfxsize=13.5cm 
\epsfbox{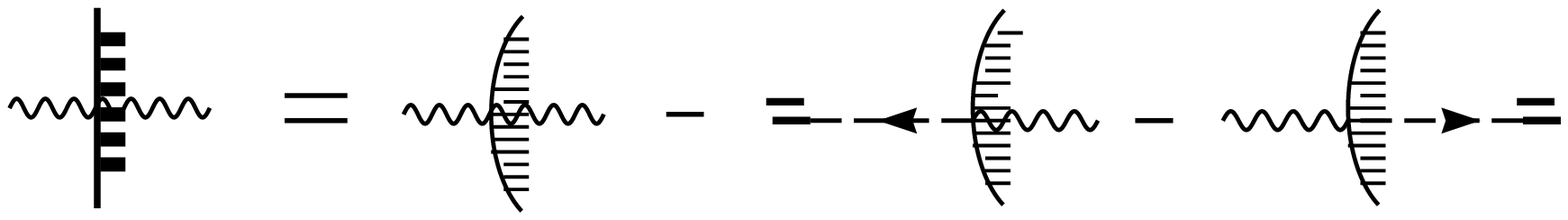}} \hfill 
\parbox{1cm}{\begin{equation}\label{R}
\end{equation}}\\ 
between cut bare propagators. Consider first  
two-particle intermediate states. Using the WI, eq.
(\ref{WI1}), repeatedly, and noting the relation 
\begin{equation}
\bar{k}_\mu k^\mu = -2(k\cdot\eta)^2 \qquad , \quad k^2=0 
\qquad ,\label{r}
\end{equation} 
one straightforwardly derives the equality \\
\parbox{14cm}{\epsfysize=3cm \epsfxsize=6cm 
\epsfbox{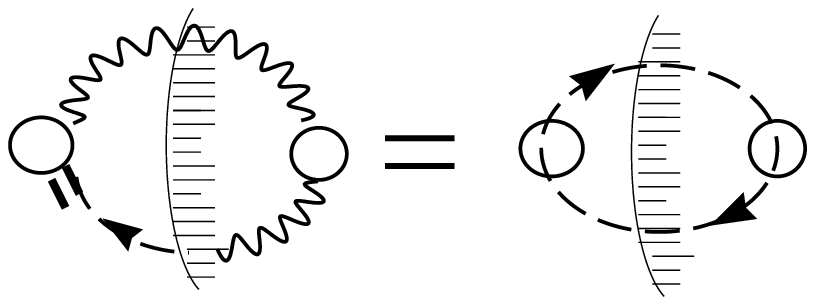}}     
\hfill \parbox{1cm}{\begin{equation}\label{Cutrel}
\end{equation}}\\ 
and hence, using also (\ref{R}), : \\
\parbox{14cm}{\epsfysize=6cm \epsfxsize=13.5cm 
\epsfbox{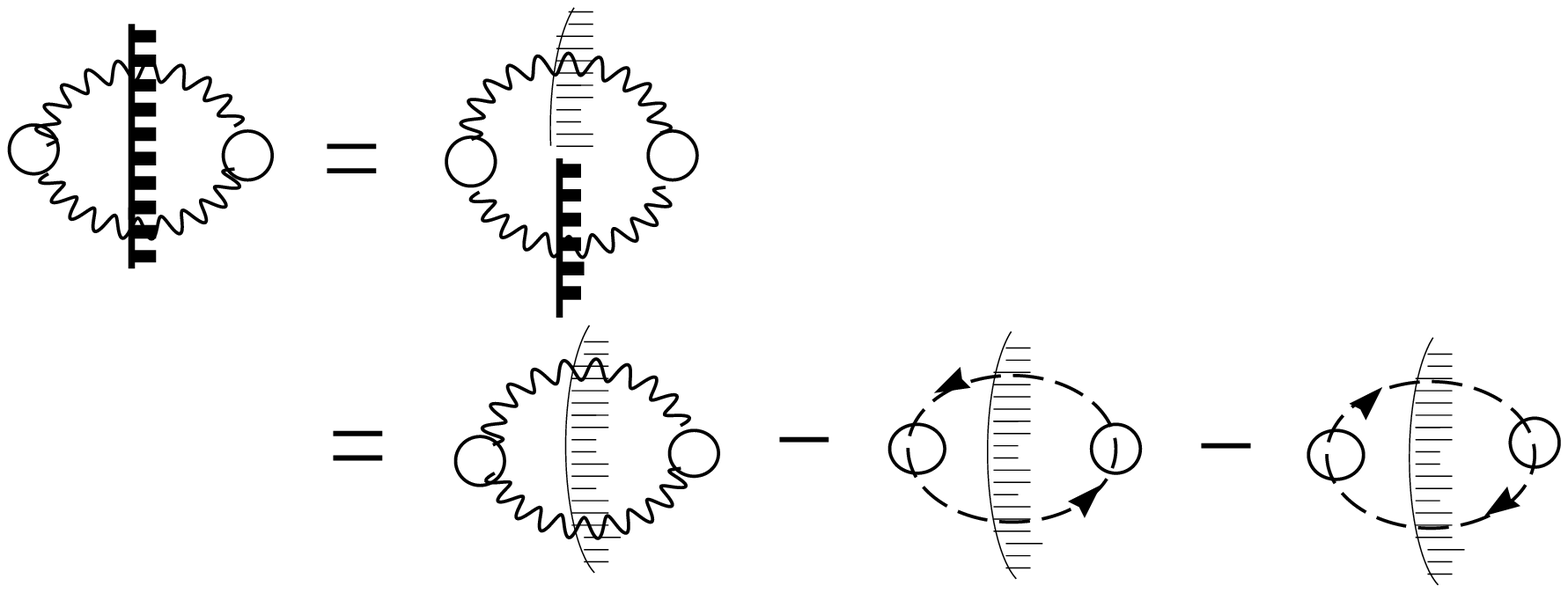}}     
\hfill \parbox{1cm}{\begin{equation}\label{PhUn}
\end{equation}}\\ 
All external legs (not shown) on the blobs in 
(\ref{Cutrel}), (\ref{PhUn}) must be on-shell physical 
gauge boson legs. (\ref{PhUn}) is the desired 
relation showing that, for two-particle intermediate 
states, the sum over cuts on the r.h.s. of 
(\ref{cuteq1}) (properly including the minus sign 
for ghost loops) reduces to the sum 
over only physical cuts (\ref{trcut}). 

The generalization to $N$-particle intermediate states 
follows directly from the same argument by means of the 
extentions of the WI (\ref{WI1}) to multiple 
external ghost legs alluded to above. Alternatively, 
the $N$-particle state can be treated by induction 
starting from the two-particle state (\ref{PhUn}). 
As, however, all this is the same as in the familiar 
local gauge theory case, there is no need to belabor 
the point. 

\section{Appendix - Cutting equations}
We remark here on some points pertaining to the 
derivation of the cutting equations for our actions. 
A Feynman graph with $N$ vertices is represented in 
coordinate space by $N$ vertex points $x_i, 
\;i=1,\ldots,N$ joined by lines. To the $i$-th 
$m_i$-point vertex factor $(2\pi)^4\,i\hat{V}_
{\mu\nu\ldots ab\ldots}(k_{i_1},\ldots,k_{i_{m_i}})
\delta(\sum_l k_{i_l})$ 
in momentum space, there corresponds the matrix 
vertex factor 
\beq
i\hat{V}_{\mu\nu\ldots ab\ldots}( x_i\,;\, x_{i_1}, 
x_{i_2}, \ldots, x_{i_{m_i}}) \equiv i\cV_{\mu\nu\ldots 
ab\ldots}(i\partial_{x_{i_1}}, \ldots, 
i\partial_{x_{i_{m_i}}})
\prod_{l=1}^{m_i}\;\delta(x_i-x_{i_l})\; .\label{coV}
\eeq
Arbitrary external source vertices are included here as 
1-point vertices. To every gauge boson (ghost) line joining 
the $l$-th leg of the vertex at $x_i$ to the $k$-th leg 
of the vertex at $x_j$ there corresponds a propagator 
factor of $D_{\mu\nu ab}(x_{i_k}-x_{j_l})\; (D_{ab}
(x_{i_k}-x_{j_l}))$. We will use a compact notation 
supressing inessential labels and letting the 
enumerative index $i$ also stand for all associated 
spacetime and group indices. Thus we denote by 
$D_{{i_k}{j_l}}$ any propagator joining 
$x_{j_l}$ to $x_{i_k}$, and by $\hat{V}(i;\, i_1\ldots
i_{m_i})$ the vertex factor at $x_i$.

Consider a graph with $N$ vertices. Associated 
with the graph is the generalized function $F(\{x_i\})$ 
defined as the product of a factor 
of $\hat{V}(i;\,i_1,\ldots,i_{m_i})$ for each vertex 
point $x_i$, and a propagator $D_{{i_k}{j_l}}$ for 
the line joining the 
$l$-th leg emanating from the vertex at $x_j$ to the 
$k$-th leg emanating at $x_i$. Note that no integration 
is included in the definition of $F(\{x_i\})$, and 
thus there is yet no distinction between internal and 
external vertices. The actual amplitude for the graph is 
obtained, apart from combinatorial factors, by 
multiplying $F(\{x_i\})$ by the appropriate external 
wave-functions, and integrating over $x_{i_k}, x_i,\;
k=1,\ldots,m_i,\;i=1,\ldots,N$. 

Given $F(x_1,\ldots,x_N)$, we define a set of related 
functions which will all be denoted by $F$ but with 
one or more of their arguments $x_i$ underlined \cite{V}. 
The function $F(x_1,\ldots,\underline{x}_i,\ldots,
\underline{x}_j,\ldots,x_N)$ with any number of 
its arguments $x_i$ underlined is defined as the product 
of the following factors:
\begin{enumerate}
\item[(a)] $i\hat{V}(i;i_1\ldots i_{m_i})$ for each $x_i$ 
that is not underlined; $-i\hat{V}(i;i_1\ldots i_{m_i})$ 
for each $x_i$ that is underlined.
\item[(b)] For a line joining the $l$-th leg of the 
vertex at $x_j$ and the $k$-th leg of the vertex at 
$x_i$:
\begin{enumerate}
\item[(i)] $D_{{i_k}{j_l}}$ if neither $x_i$ nor $x_j$ 
are underlined.
\item[(ii)] $D^*_{{i_k}{j_l}}$if both $x_i$ and $x_j$ 
are underlined. 
\item[(iii)] $D^+_{{i_k}{j_l}}$ if $x_i$ but not $x_j$ 
is underlined. 
\item[(iv)] $D_{{i_k}{j_l}}^-$ if $x_j$ but not $x_i$ 
is underlined. 
\end{enumerate}
\end{enumerate}
Note that (iii) implies (iv), and vice-versa, because of 
the relation $D^+(-x))=D^-(x)$. This makes the inherent 
ambiguity in assigning a direction to propagators, which 
are symmetric in their arguments, irrelevant; the point is 
that positive energy always flows from the not underlined 
to the underlined vertex, whereas there is no restriction 
on the sign of energy flow for lines connecting two 
vertices which are both either underlined or not 
underlined. Underlining vertices is clearly related to 
complex  conjugation provided all vertex factors 
$\hat{V}(i;i_1\ldots i_{m_i})$ are real. Given 
a graph with some vertices underlined, consider the graph 
obtained by removing all present underlinings, and 
underlining all previously not underlined vertices. It 
is easily verified that the 
respective functions $F$ are complex conjugate of each 
other.

Each vertex $\hat{V}_{\mu\nu\ldots ab\ldots}(k_{i_1},
\ldots,k_{i_{m_i}})$ can be 
represented by an everywhere absolutely convergent 
series. By standard theorems, the series of generalized 
functions obtained by Fourier transforming converges 
to the corresponding Fourier transform (\ref{coV}). Orders 
of summations may be interchanged, and the essentially 
combinatorial argument \cite{V} on the functions 
$F(\{x_i\})$ leading to the largest time and cutting 
equations may in fact be applied term by term in the 
series representations. Infinite radius 
of convergence is crucial for this argument to be 
applicable in the present case. Finite radius 
corresponding to singularities in the vertices would mean 
that the  argument either fails or must be modified in a 
manner resulting into cut contributions from vertices. 
The other necessary ingredients, also required in the 
usual polynomial action case, are: (a) the quasi-local 
nature of each (term in the expansion of) vertex  
$\hat{V}(i;\, i_1\ldots i_{m_i})$ (i.e. proportional 
to delta functions of $(x_i-x_{i_k})$ and their 
derivatives); (b) the propagators $D_{{i_k}{j_l}}$ 
obeying the decomposition (\ref{Fpdec}).

Assuming then a $x_j^0$ to be the largest among the 
time components of the points $\{x_i|i=1,\ldots,N\}$, 
the argument of \cite{V} may now be applied to give 
a largest time equation, and hence, by summation over 
orderings, the result: 
\beq
F(\{x_i\}) + F(\{x_i\})^* = - \sum_{\{\,\underline{\ }\, 
\backslash 0,\;all\}} F(\{x_i\})  \label{rS}
\eeq   
In (\ref{rS}) the sum is over the $(2^N-2)$ possible 
underlinings other than no and all underlinings. 
Multiplying (\ref{rS}) by any appropriate external 
wave functions and integrating over all 
$\{x_{i_k}\},x_i$ gives (\ref{cuteq}), the cutting 
equation (in momentum space). By noting that many terms on 
the r.h.s. vanish due to conflicting energy 
$\theta$-functions, it is easily seen that the sum over 
underlinings indeed reduces to the sum over cuts as 
described in the main text. 

The following point should be noted. Although (\ref{rS}) 
holds for any time ordering of the $x_i$'s (thus allowing 
one to integrate), it was established, for each given 
configuration $\{x_i\}$, only for one of the time components 
$x_i^0$ being the largest, i.e. not for the case of two or 
more of the time components being equal and largest than the 
rest. Such equal time regions are of lower dimensionality, 
hence measure zero, in the $N$-dimensional integration 
space of the $x_i$'s. They therefore give no finite 
contribution provided the (regulated or subtracted) 
Feynman integrands are sufficiently regular in 
coordinate space.\footnote{A similar proviso applies to 
tadpoles, formed by lines closing upon themselves, though 
these actually give no contribution to the sum over cuts. 
The use of dimensional (or equivalent) regularization which 
sets infinite constants such as $\delta^{(n)}(0), \;\;n\geq0,$ 
to zero is extremely convenient in automatically handling 
these subtleties.}

Return to (\ref{rS}), and assume that $x_k^0 < x_l^0$. Then 
the equation holds separately for the terms with and 
without $x_k$ being underlined since it is certain that 
$x_k$ is not the largest time \cite{V}. So, in particular, 
\beq
\sum_{\{\underline{\ }\backslash k\}} F(\{x_i\}) = 0 
\quad,\qquad  x_k^0 < x_l^0 \quad , \label{cS}
\eeq
where the sum is over all underlinings except $x_k$. 
Similarly, considering the case $x_k^0> x_l^0$ gives 
equation (\ref{cS}) with $x_k$ and $x_l$ interchanged. 
Adding these two equations gives the relation: 
\bea
F(\{x_i\}) = - \sum_{\{\,\underline{\ }\,\backslash k,l\}} 
F(\{x_i\})& - & \theta(x^0_l - x_k^0) \sum_{l\,\cup\{\,
\underline{\ }\backslash k,l\}} F(\{x_i\}) \nonumber\\
   & - & \theta(x^0_k - x_l^0) \sum_{k\,\cup\{
\,\underline{\ }\backslash k,l\}} F(\{x_i\})\ , \;\; 
x_k^0 \neq x_l^0 .\label{bccS}
\eea
In (\ref{bccS}) the term with no underlining and the term 
with neither $x_k$ nor $x_l$ underlined have been separated 
out. The two terms multiplied by the $\theta$-functions 
contain the sums of all underlinings with $x_l$ but not 
$x_k$ underlined, and with $x_k$ but not $x_l$ underlined, 
respectively.  

Multiplying (\ref{cS}) by any appropriate wave-functions 
attached to external vertices, and integrating over all 
$x_{i_k}, k=1,\ldots,m_i,i=1,\ldots,N$, and all $x_i$ 
except $x_k, x_l$ gives the BCC equation as originally 
stated \cite{BS}.
An equivalent and more convenient statement \cite{V} 
is obtained by performing the same operations on 
(\ref{bccS}) which results in (\ref{bcc}) in the main 
text with $x_k=x, x_l=y$. 

Finally, integrating also over $x, y$ allows one to 
express the BCC equation (\ref{bcc}), entirely in 
momentum space, in the form of a dispersion relation. 
In performing this step, however, one must note that 
(\ref{bccS}) was derived strictly 
only for $x^0\neq y^0$. In the presence of derivative 
interactions, finite contributions can arise from 
the point $x^0=y^0$ due to the action of derivatives 
at $x,y$ on the $\theta$-functions in (\ref{bcc}). 
The general rule is as follows. Any derivatives  
which, upon Fourier-transforming, correspond to 
external momenta injected into the graph along external 
lines (truncated vertex legs) attached at $x$ and/or 
$y$, or along a propagator leg connecting a source vertex 
at $x$ and/or $y$ to the rest of the diagram, must also 
act on the $\theta(\pm(x^0-y^0))$ factors, i.e. must be 
commuted to the front on the r.h.s. in (\ref{bcc}) (cp. 
\cite{BS}). 

As an example, consider the important case of the 
$2$-point function between external sources $J_x,\; J_y$ 
at $x$ and $y$. To lowest order  where the blob  
stands for a single bare propagator joining $x$ and 
$y$, (\ref{bcc}) is nothing but (\ref{pdec}), in 
Feynman gauge, sandwiched between $J_x^{\mu a}$ and 
$J_y^{\nu b}$. Comparison with (\ref{cop}) trivially 
shows that the correct extension to include the point 
$x^0=y^0$ is obtained by letting derivatives in 
$\hat{J}_x=J_x\oh^{\;-1/2}(-\Box_x/\Lambda^2)$ and 
$\hat{J}_y$ also act on $\theta(\pm(x^0-y^0))$. Similarly, 
in the presence of an arbitrary number $n$ of 
self-energy insertions between $x$ and $y$, one has: 
\bea
 & &\int\:\prod_{k=i}^n \;dz_k\; J_x\cdot\oh^{\;-1/2}
(-\Box_x)\cdot D_{x{z_1}}\cdot\oh^{\;-1/2}(-\Box_{z_1})
\cdot\Pi_{{z_1}{z_2}}\cdot\oh^{\;-1/2}(-\Box_{z_2})
\cdot D_{{z_2}{z_3}}\nonumber\\
   & & \qquad\qquad \cdots \Pi_{{z_{n-1}{z_n}}}\cdot
\oh^{\;-1/2}(-\Box_{z_n})\cdot D_{{z_n}{z_y}}
\cdot\oh^{\;-1/2}(-\Box_y)\cdot J_y\nonumber\\
  & = & J_x\cdot\oh^{\;-(n+1)}(-\Box_x)\cdot\left[ 
\int\!\prod_{k=i}^n dz_k\: D_{x{z_1}}
\cdot\Pi_{{z_1}{z_2}}
\cdot D_{{z_2}{z_3}}\cdots \Pi_{{z_{n-1}{z_n}}}
\cdot D_{{z_n}{z_y}}\right]\cdot J_y \label{2S} 
\eea 
by integration by parts and translation invariance 
of $D_{z{z^\prime}},\;\Pi_{z{z^\prime}}$.   
Again, in applying (\ref{bcc}) to (\ref{2S}), all 
derivatives in sources and the $n+1 \;\oh^{\;-1}$ factors 
must also act on the $\theta$-factors on the r.h.s. 
This clearly generalizes to the propagator leg 
connecting a source vertex at $x$ and/or $y$
to any $(2+n)$-legged blobs in (\ref{bcc}).

\end{document}